\documentclass[12pt]{article}
\usepackage{arxiv}
\usepackage{natbib}
\usepackage{url}
\usepackage{amsmath, bm}
\usepackage{amsfonts}
\usepackage{caption}
\usepackage{subcaption}
\usepackage{graphicx,psfrag,epsf}
\usepackage{enumerate}
\usepackage{longtable}
\usepackage{anyfontsize}
\usepackage{booktabs}
\usepackage{float}
\usepackage{xcolor}
\usepackage{xr}
\usepackage[flushleft]{threeparttable}
\usepackage{changepage}
\newcommand{\given}{\,|\,}

\newcommand{\bA}{\mathbf{A}}

\newcommand{\calV}{{\cal V}}
\makeatletter
\newcommand*{\addFileDependency}[1]{
	\typeout{(#1)}
	\@addtofilelist{#1}
	\IfFileExists{#1}{}{\typeout{No file #1.}}
}
\makeatother

\newcommand*{\myexternaldocument}[1]{%
	\externaldocument{#1}%
	\addFileDependency{#1.tex}%
	\addFileDependency{#1.aux}%
}
\myexternaldocument{supinfo_revision}

\begin{document}
\title{Bayesian Models for Multivariate Difference Boundary Detection in Areal Data}

\author{LEIWEN GAO, SUDIPTO BANERJEE*\\[4pt]
	\textit{Department of Biostatistics, University of California, Los Angeles, Los Angeles, California, 90095, USA}
	\\[2pt]
	{sudipto@ucla.edu}\\ 
	BEATE RITZ\\[4pt]
	\textit{Department of Epidemiology, University of California, Los Angeles, Los Angeles, California, 90095, USA}\\[2pt]
}

\markboth%
{L. Gao and others}
{}

\maketitle

\footnotetext{To whom correspondence should be addressed.}

\begin{abstract}
	{Regional aggregates of health outcomes over delineated administrative units (e.g., states, counties, zip codes), or areal units, are widely used by epidemiologists to map mortality or incidence rates and capture geographic variation. To capture health disparities over regions, we seek ``difference boundaries'' that separate neighboring regions with significantly different spatial effects. Matters are more challenging with multiple outcomes over each  unit, where we capture dependence among diseases as well as across the areal units. Here, we address multivariate difference boundary detection for correlated diseases. We formulate the problem in terms of Bayesian pairwise multiple comparisons and seek the posterior probabilities of neighboring spatial effects being different. To achieve this, we endow the spatial random effects with a discrete probability law using a class of multivariate areally-referenced Dirichlet process (MARDP) models that accommodate spatial and inter-disease dependence. We evaluate our method through simulation studies and detect difference boundaries for multiple cancers using data from the Surveillance, Epidemiology, and End Results (SEER) Program of the National Cancer Institute.}
	{Disease mapping; Dirichlet process; Directed acyclic graphical autoregression; False discovery rates; Multivariate spatially dependent models; Wombling.}
\end{abstract}

\section{Introduction}
\label{sec:1}

Spatial data analysis in public health applications often proceed from statistical models for \emph{areal data} that comprises regional aggregates of health outcomes over delineated administrative units such as states, counties or zip codes. Disease mapping, in particular, is an epidemiologic exercise that models spatial dependence of counts or rates (e.g., incidence or mortality) to better understand geographic variation of diseases \citep{koch2005cartographies, lawson2016handbook}. Spatial dependence is introduced using stochastic models on graphs, where the nodes correspond to regions and an edge between two nodes relate them as neighbors. Examples include Markov random fields using undirected graphs \citep{rueheld04, besag1974spatial, besag1991bayesian, kissling2008spatial} or directed acyclic graphical autoregression (DAGAR) models \citep{datta2018spatial}. 

In this article we address one important aspect of disease mapping: identifying \emph{difference boundaries} that separate regions with significantly different spatial random effects from their neighbors. 
This exercise has sometimes been referred to as \emph{areal wombling} \citep[named after a seminal paper by][]{womble1951differential} in spatial data science, including spatiotemporal boundary analysis 
{\citep{berchuck2019jasa}}, but has largely been restricted to a single outcome \citep{li2011mining, jacquez2003local, lu2005bayesian, lu2007bayesian, ma2010hierarchical,  jacquez2003geographic} with the notable exceptions of \cite{carlin2007bayesian}, who implemented a deterministic algorithm to compare posterior estimates from multivariate CAR (MCAR) models, 
{and \cite{corpas-burgos2020serra} who used adaptive spatial weights in multivariate models}. However, these approaches may have limited capabilities for probabilistic inference or for propagating uncertainty estimates on the difference boundaries.   

In this article we formulate this problem in terms of Bayesian multiple comparisons, where we evaluate the posterior probability that the spatial random effects from a pair of adjacent regions are different. These posterior probabilities are computed for all pairwise adjacencies on the map and subsequently controlled using (Bayesian) False Discovery Rates (FDR) \citep{muller2004optimal}. Given that we are evaluating the posterior probabilities of the equality (or not) of spatial random effects, we must endow the spatial effects with a discrete probability law. 

Due to associations emanating from a shared set of unaccounted factors such as genetic and environmental risks, therapeutic success, or early diagnosis  the presence of one disease can aggravate (or inhibit) the occurrence of others in the same or neighboring regions. This generates associations among diseases \citep[see, e.g.,][]{lindstrom2017quantifying, agrawal2018risk, shi2004frequencies}. Multivariate areal models for continuous random effects \citep[see, e.g., a comprehensive discussion by][and references therein]{mcnab2018} have demonstrated the statistical benefits of jointly modeling multiple diseases across areal units. Fitting independent univariate models for each disease yields biases from ignoring dependencies among diseases. Joint models are often constructed using multivariate Markov random fields \citep{mardia1988multi}, although alternatives using Moran's I basis functions have also been developed \citep{bradley2015multivariate, bradley2018computationally}. 

For discrete multivariate spatial distributions, one could build upon classes of parametric univariate discrete spatial moving average models (SMA) \citep[see, e.g.][]{banerjee2012bayesian}. However, inference from such models is sensitive to prior specifications. Instead, we expand upon a demonstrably effective nonparametric approach for univariate boundary detection proposed by \citet{li2015bayesian} and further elucidated by \citet{hanson2015spatial} and extend them to analyze multiple correlated diseases. More specifically, we achieve probabilistic estimation for difference boundaries by embedding a multivariate areal model within a hierarchical Dirichlet process model. We call this a Multivariate Areal Dirichlet Process (MARDP). 

Others have adopted different viewpoints on boundary detection from ours. For example, \citet{qu2021boundary} proposed an integrated stochastic process to infer on boundaries based upon continuous gradients as defined in curvilinear wombling \citep{banerjee2006bayesian}. While attractive for continuous random fields where a ``wombling boundary'' is defined as one located in a zone with high directional gradients, our difference boundaries are a subset of administrative boundaries defined on the basis of significantly different spatial effects. \citet{ma2010hierarchical} used a stochastic edge mixed effects (SEME) model for unknown adjacencies and detected the presence of edges by incorporating covariates. The detection of edges was only used for the improvement of spatial effects estimation but not difference boundary detection. 
{Estimating adjacencies in areal modeling contexts include \cite{lu2007bayesian}, \cite{lee2021statcomp} and \cite{corpas-burgos2020serra}, while other related approaches include modeling discontinuities \citep[][]{santafe2021smmr} and step changes \citep[][]{rushworth2017jrssc} in disease risk.}

Turning to FDR-based methods, we note the work by \citet{perone2004false} for testing an uncountable set of hypothesis tests on Gaussian random fields and the FDR smoothing developed by \citet{tansey2018false} that exploits spatial structure within a multiple-testing problem. The former pertains to point-referenced data, while we focus on areal data. The latter focused on identifying regions with enriched local fraction of signals against the background, while we intend to ascertain difference boundaries based upon the differences between latent spatial random effects after accounting for risk factors, confounders and other explanatory variables.

The paper is organized as follows. Section~\ref{methods} develops our MARDP framework for discrete spatial random effects comprising a multivariate DAGAR model and an FDR-based rule for multivariate areal boundary analysis. Section~\ref{sim} presents a simulation study to assess the performance of different models, while Section~\ref{data_analysis} conducts boundary analysis on a multivariate areal dataset for standardized incidence ratios (SIR) of four cancers in California obtained from SEER.

\section{Methods}
\label{methods}
Modeling multiple diseases will introduce associations among the diseases as well as spatial dependence for each disease. For $q$ diseases, let $y_{id}$ denote a disease outcome of interest for disease $d$ in region $i$ such that $d = 1, \dots, q$, $i = 1, \dots, n$. 
{We assume a typical generalized linear mixed model setting where $y_{id}$ follows a distribution from the exponential family 
with %
canonical link
\begin{equation}
g(\mbox{E}(y_{id})) = \bm{x}_{id}^\top\bm{\beta}_d + \phi_{id}
\label{eq: spatial_regression}
\end{equation}}where $\bm{x}_{id}$ is a $p_d\times 1$ vector of explanatory variables specific to disease $d$ within region $i$, $\bm{\beta}_d$ are the slopes corresponding to disease $d$, %
and 
$\phi_{id}$ is a random effect for disease $d$ in region $i$
. 
In 
Section~\ref{data_analysis} we analyze count data using hierarchical Poisson regression.

Part of the residual in (\ref{eq: spatial_regression}) is captured by the spatial random effect $\phi_{id}$ for disease $d$ in region $i$. For boundary detection, we define difference boundaries by considering probabilities such as $P(\phi_{id} = \phi_{jd} | i \sim j)$ and $P(\phi_{id} = \phi_{jd'} | i \sim j, d \neq d')$, where $\sim$ denotes spatial neighbors. If the $\phi_{id}$'s are continuous, the probabilities will always be 0 which do not work for boundary detection. Instead, we build a multivariate areal Dirichlet process (MARDP) that accommodates spatial dependence while modeling spatial random effects as discrete variables. With more than one disease of interest, we also introduce associations among diseases within our framework.

\subsection{The Multivariate Areally Referenced Dirichlet Process}\label{ARDP}
We extend the univariate modeling framework in \citet{li2015bayesian} to a multivariate model for $q \geq 2$.  Let $N = n \times q$ be the total number of observations and let $\bm{\phi} = \left(\bm{\phi}_{1}^\top,\dots,\bm{\phi}_{q}^\top\right)^{\top}$, where $\bm{\phi}_{d} = \left(\phi_{1d}, \dots, \phi_{nd}\right)^\top$. Let $(1,1), \dots, (n,1), (1,2), \dots, (n,2),\dots, (1, q), \dots, (n,q)$ be the pairwise $(i, d)$ indices corresponding to a vectorized enumeration of the observations $1, \dots, n, n+1, \dots, 2n, \dots, (q-1)n+1, \dots, N$. For $\bm{\theta} = \left(\theta_1, \dots, \theta_K\right)$, each $\theta_k$, $k=1,\dots, K$, is a random sample drawn independently from a base distribution $N(0, 1/\tau_s)$ with precision $\tau_s$. Letting $\delta_{\theta_k}$ be the Dirac measure located at $\theta_k$ and modeling $\bm{\phi}$ jointly as an unknown distribution $G_N$, which itself is modeled as a Dirichlet process (DP), yields the Multivariate Areal DP (MARDP)
\begin{align}
&\bm{\phi} \sim G_{N}\;;\quad G_{N} | \pi_{u_1, \dots, u_N}, \bm{\theta} = \sum_{u_1, \dots, u_N}\pi_{u_1, \dots, u_N}\delta_{\theta_{u_1}}\dots\delta_{\theta_{u_N}};\nonumber\\ 
&\pi_{u_1, \dots, u_N} = Pr\left(\sum_{k=1}^{u_1-1}p_k < F^{(1)}(\gamma_1) < \sum_{k=1}^{u_1}p_k,\dots,\sum_{k=1}^{u_N-1}p_k < F^{(N)}(\gamma_N) < \sum_{k=1}^{u_N}p_k\right); \nonumber\\
&\bm{\gamma} = \left\{\bm{\gamma}_1, \dots, \bm{\gamma}_q\right\}\sim N_{nq}(\bm{0}, \bm{\Sigma}_\gamma) \label{eq:DP}
\end{align} 
where $p_1, \dots, p_K$ correspond to the stick breaking weights \citep{sethuraman1994constructive} constructed as $p_{1} = V_1$ and $p_{j} = V_j\prod_{k<j}(1-V_k)$ for each $j = 2, \dots, K$, where each $V_k \overset{\text{iid}}{\sim} Beta(1,\alpha)$
and $u_1, \dots, u_N$ are indices of $\theta_k$'s sampled for the $N$ observations. The total number of DP clusters, $K$, truncates the stick breaking function. The infinite sum of $p_k$'s is 1. Spatial components $\bm{\gamma}_d = (\gamma_{1d}, \gamma_{2d}\dots, \gamma_{nd})^\top$ are dependent for each disease $d$, and are modeled jointly with covariance matrix $\bm{\Sigma}_\gamma$. Each $F^{(1)}(\cdot), \dots, F^{(N)}(\cdot)$ (corresponding to $F^{(1,1)}(\cdot), \dots, F^{(n,q)}(\cdot)$, respectively) denotes the cumulative distribution functions of the marginal distribution of the corresponding $\gamma_{id}$. Marginally, each $F^{(i,d)}(\gamma_{id}) \sim Uniform(0,1)$ but dependence is introduced through the $\gamma_{id}$'s. 
The marginal distribution for the individual $\phi_{id}$ is given as $G^{(i,d)}(\phi_{id}) = \sum_{k=1}^K\pi_k\delta_{\theta_k}$, where $\pi_k = P\left(\sum_{t=1}^{k-1}p_t < F^{(i,d)}(\gamma_{id}) < \sum_{t=1}^{k}p_t\right)$. These DPs are dependent across regions as well as diseases with dependent $F^{(i,d)}(\gamma_{id})$'s and through $p_1, \dots, p_K$. Hence, the MARDP framework is able to evaluate the difference in $\phi_{id}$'s across diseases. The shared values of the $\theta_k$'s enable comparisons of spatial effects between diseases. We next turn to $\bm{\Sigma}_{\gamma}$.


\subsection{Joint Multivariate DAGAR Model for Spatial Components} \label{MJDAGAR}
The hierarchical MARDP framework depends upon a valid (positive definite) choice for $\bm{\Sigma}_{\gamma}$. Covariance matrices from ``proper'' MCAR models \citep[e.g.,][]{gelfand2003proper, sain2007, mcnab2016} present 
such %
choices
. Inferential benefits of univariate DAGAR \citep[][]{datta2018spatial} for spatial autocorrelation 
motivates %
Multivariate DAGAR (MDAGAR).  

Following \citet{jin2007order} we let $\bm{\gamma}_d$ be a linear combination of latent factors $\bm{f}_1, \dots, \bm{f}_d$ for $d \geq 2$, where each $\bm{f}_d \sim N(0, \bm{Q}_d^{-1})$ is independently modeled as a univariate DAGAR. DAGAR uses any fixed ordered set of regions, $\calV = \{1,2,\ldots,n\}$ to construct geographic neighbors of $i$, say $N(i)$, comprising regions that \emph{precede} $i$ in $\calV$. The precision matrix is constructed as $\bm{Q}(\rho) = (\bm{I}-\bm{B})^{\top}\bm{\Lambda}(\bm{I}-\bm{B})$, where $\bm{B}$ is a $n\times n$ strictly lower-triangular matrix with elements $b_{ij}=0$ if $j\notin N(i)$ and $\displaystyle b_{ij} = \frac{\rho}{1 + (n_{<i}-1)\rho^2}$ for $i=2,\ldots,k$ and $j\in N(i)$; and $\bm{\Lambda}$ is a $n\times n$ diagonal matrix with elements $\displaystyle \lambda_{i} = \frac{1 + (n_{<i}-1)\rho^2}{1-\rho^2}$ for $i=1,2,\ldots,k$ with $n_{<i}$ being the number of members in $N(i)$ and $n_{<1} = 0$. \cite{datta2018spatial} show that $\rho$ acts as an easily interpretable spatial autocorrelation parameter in the above graphical autoregression structure.
 
With $\bm{f}_d \stackrel{ind}{\sim} N(\bm{0}, \bm{Q}_d^{-1})$, where $\bm{Q}_d = \bm{Q}(\rho_d)$, the joint distribution for $\bm{\gamma}$ is constructed from $\bm{\gamma}_1 = a_{11}\bm{f}_1$ and $\bm{\gamma}_d = a_{d1}\bm{f}_1 + a_{d2}\bm{f}_2 + \dots + a_{dd}\bm{f}_d$ for each $d = 2, \dots, q$,
where $a_{dh}, h = 1, \dots, d$, are coefficients that associate spatial components for different diseases. If $\bm{F} = (\bm{f}_1^\top, \dots, \bm{f}_q^\top)^\top$ and $\bm{A}$ is the lower triangular matrix with elements $a_{dh}$, then the covariance matrix of $\bm{\gamma}$ is
\begin{align}\label{eq:Sigma_g}
\bm{\Sigma}_\gamma &= \left(\bm{A}\otimes\bm{I}_k\right)Cov(\bm{F})(\bm{A}\otimes\bm{I}_k)^\top = (\bm{A}\otimes\bm{I}_k) \left[\bigoplus_{d=1}^q \bm{Q}^{-1}(\rho_d)\right] \left(\bm{A}^\top\otimes\bm{I}_k\right)\;.
\end{align}
With a shared $\rho_d=\rho$ for all diseases, we obtain a separable covariance matrix $\bm{\Sigma}_\gamma = (\bm{A}\bm{A}^\top)\otimes \bm{Q}^{-1}(\rho)$ as the Kronecker product of $\bm{A}\bm{A}^\top$ which corresponds to disease dependence and $\bm{Q}^{-1}(\rho)$ corresponding to spatial association. Henceforth, when $\bm{\Sigma}_\gamma$ is defined as in \eqref{eq:Sigma_g}, we will refer to the MARDP framework in \eqref{eq:DP} simply as MDAGAR($\rho_1, \dots, \rho_q, \bm{\Sigma}$), while if $\bm{Q}_d$ is specified using a proper CAR structure, i.e. $\bm{Q}_d = \bm{D} - \rho_d\bm{M}$, where $\bm{D}$ is diagonal with $D_{ii}$ (i.e. the number of neighbors of region $i$) and $\bm{M}$ is the binary adjacency matrix for the map ($M_{ii} = 0$, $M_{ij} = 1$ if $i \sim j$ and $M_{ij} = 0$ otherwise), then we will refer to \eqref{eq:DP} simply as MCAR($\rho_1, \dots, \rho_q, \bm{\Sigma}$), i.e. the order-free multivariate CAR model proposed by \citet{jin2007order}. 
{The MDAGAR and MCAR models so constructed are not specific to ordering of diseases, unlike \citet{gao2021hierarchical} or \cite{jin2005generalized}, hence are applicable to more than a few diseases.}

\subsection{Model Implementation}
\label{implementation}
We extend \eqref{eq: spatial_regression} to a Bayesian hierarchical framework with the posterior distribution
\begin{align}\label{eq:post_dist}
p\left(\bm{\beta}, \bm{\phi}, \bm{\theta}, \bm{\gamma}, \bm{V}, \bm{\tau}, \tau_s, \bm{\rho}, \bm{A} \given \bm{y}\right) \propto p\left(\bm{\beta}, \bm{\phi}, \bm{\theta}, \bm{\gamma}, \bm{V}, \bm{\tau}, \tau_s, \bm{\rho}, \bm{A}\right) \times \prod_{d=1}^q\prod_{i=1}^n 
{p(y_{id}\given \bm{\beta}_d, \phi_{id}, {\tau}_d)}
\end{align}
where 
{$p(y_{id}\given \bm{\beta}_d, \phi_{id}, \tau_d)$ is 
specified by $\{\bm{\beta}_d, \phi_{id}, {\tau}_d\}$. For example, if $y_{id}\stackrel{ind}{\sim} N(\bm{x}_{id}^{\top}\bm{\beta}_d + \phi_{d}, 1/\tau_d)$, where $\tau_d$ is the precision, then $p\left(\bm{\beta}, \bm{\phi}, \bm{\theta}, \bm{\gamma}, \bm{V}, \bm{\tau}, \tau_s, \bm{\rho}, \bm{A}\right)$ can be specified as
\begin{align}\label{eq: prior}
& \prod_{k=1}^K\left\{N(\theta_k\given 0,1/ \tau_s)\times Beta(V_k\given 1, \alpha)\right\}\times \prod_{d=1}^q \left\{IG(1/\tau_d\given a_{e}, b_{e}) \times N(\bm{\beta}_d\given \bm{0}, \sigma_\beta^2\bm{I}_{p_d}) \times Unif(\rho_d \given 0,1) \right\} \nonumber\\
&\quad \times IG(1/\tau_s \given a_{s}, b_{s}) \times N\left(\bm{\gamma} \given \bm{0}, \bm{\Sigma}_\gamma(\bm{\rho}, \bm{A})\right) \times IW\left(\bm{A}\bm{A}^{\top} \given \nu, \bm{R}\right)\times \left|\frac{\partial \bm{\Sigma}}{\partial a_{dh}}\right|\;,
\end{align}
where $\bm{\tau} = \{\tau_1,\ldots,\tau_q\}$,} $\left|\frac{\partial \bm{\Sigma}}{\partial a_{dh}}\right|$ is the Jacobian $2^q\prod_{d=1}^qa_{dd}^{q-d+1}$ transformation for the prior on $\bA\bA^{\top}$ in terms of the Cholesky factor $\bA$. We sample the parameters from the posterior distribution in \eqref{eq:post_dist} using Markov chain Monte Carlo (MCMC) with Gibbs sampling and random walk metropolis \citep{gamerman2006markov} implemented in the \texttt{R} statistical computing environment. Section ~\ref{algorithm} presents details on the MCMC updating scheme. 

\subsection{Decision Rule Based on FDR for Selecting Difference Boundaries}

Following \cite{li2015bayesian} we formulate difference boundary detection as a multiple comparison problem, where a cancer-specific difference boundary is detected according to the tenability, or not, of $\phi_{id} = \phi_{jd}$ for $i \sim j$. To adjust for the multiplicity arising from all pairs of neighbors and, in our case, of diseases as well, a false discovery rate (FDR) is controlled \citep{benjamini1995controlling}. We adopt the Bayesian analogue of FDR \citet{muller2004optimal} in the following manner: We define an edge $(i, j)^d$ as a difference boundary for disease $d$ if $P(\phi_{id} \neq \phi_{jd} \given \bm{y})$ exceeds a certain threshold $t$. Denoting $v_{(i,j)}^d = P(\phi_{id} \neq \phi_{jd} \given \bm{y})$, we define $\displaystyle FDR = \frac{\sum_{i\sim j} I\left(\phi_{id} = \phi_{jd}\right)I\left(v_{(i,j)}^d>t\right)}{\sum_{i\sim j}I\left(v_{(i,j)}^d>t\right)}$, and the estimated FDR is obtained as the posterior expectation
\begin{align}
\overline{FDR} = \frac{\sum_{i\sim j} \left(1-v_{(i,j)}^d\right)I\left(v_{(i,j)}^d>t\right)}{\sum_{i\sim j}I\left(v_{(i,j)}^d>t\right)}. \label{eq:est_FDR}
\end{align}
We also compute $\displaystyle \overline{FNR} = \frac{\sum_{i\sim j} v_{(i,j)}^d\left(1-I\left(v_{(i,j)}^d>t\right)\right)}{m-\sum_{i\sim j}I\left(v_{(i,j)}^d>t\right)}$ to estimate the False Non-discovery Rate (FNR), where $m$ is the total number of edges (geographic boundaries). In terms of a bivariate loss function $L_{2R} = \left(\overline{FDR}, \overline{FNR}\right)$, the optimal decision minimizes $\overline{FNR}$ subject to $\overline{FDR} \leq \delta$, i.e. the threshold $t = t^\star$ is obtained as \citep{muller2004optimal}:
\begin{align}
t^\star = \text{sup}\left\{t: \overline{FDR}(t) \leq \delta \right\}\;.\label{eq: threshold}
\end{align} 

The posterior probability $v_{(i,j)}^d$ in \eqref{eq:est_FDR} is defined according to the type of difference boundary. For instance, we use $v_{(i,j)}^s = P\left(\phi_{id} \neq \phi_{jd},  \phi_{id'} \neq \phi_{jd'}\given \bm{y}\right)$ for shared boundaries and $v_{(i,j)}^c = P\left(\phi_{id} \neq \phi_{jd'}, \phi_{id'} \neq \phi_{jd} | \bm{y}\right), i < j$ for mutual cross-disease difference boundaries ($d$ and $d'$ are two different diseases) instead of $v_{(i,j)}^d$ in \eqref{eq:est_FDR}.

\section{Simulation}
   \label{sim}
   We present a simulation experiment to compare the performances of MDAGAR and MCAR with two independent-disease models, 
   {as well as an existing multivariate method, the MCAR-based boundary likelihood values (MBLV) approach \citep{carlin2007bayesian}}. All models were constructed using the MARDP framework in Section~\ref{ARDP} and differ only in their specification of $\bm{\Sigma_{\gamma}}$. 
   
   \subsection{Data Generation}
   We generate data over a California county map with 58 counties. 
   We simulated our outcomes 
   {$y_{id}\stackrel{ind}{\sim} N(\bm{x}_{id}^{\top}\bm{\beta}_d + \phi_{id}, 1/\tau_d)$} 
   with $q=2$, i.e., two outcomes, and two covariates, $\bm{x}_{i1} = (1, x_{i12})^\top$ and $\bm{x}_{i2} = (1, x_{i22})^\top$, with $p_1=p_2=2$. We fixed values of $x_{i12}$ and $x_{i22}$ by generating them from $N(0, 1)$ independently across regions. The regression slopes were fixed at $\bm{\beta}_1 = (2, 5)^{\top}$ and $\bm{\beta}_2 = (1,6)^{\top}$ and $\tau_1 = \tau_2 = 10$. For the spatial effects, we generated values of $\bm{\phi} = \left(\bm{\phi}_1^{\top}, \bm{\phi}_2^{\top}\right)^{\top}$ using (\ref{eq:DP}) with $K=15$, $\alpha=1$, and $\tau_s = 0.25$, while we generated values for $\bm{\gamma}$ from $N(\bm{0},\bm{\Sigma}_\gamma)$ with $\bm{\Sigma}_\gamma$ in (\ref{eq:Sigma_g}) specified by $\displaystyle \bm{A} = \begin{pmatrix}1 & 0 \\ 1 & 1\end{pmatrix}$
   , $\bm{Q}^{-1}(\rho_d)$ is a spatial autocorrelation matrix with elements $\rho_d^{d(i,j)}$, $\rho_1 = 0.2$ and $\rho_2 = 0.8$, where $d(i,j)$ refers to the distance between the centroids of the $i$th and $j$th counties in California. 
   {This setup offers a ``neutral'' ground to compare MDAGAR with MCAR since the spatial structure corresponds to a covariance function based upon point-referenced centroids of regions, rather than areal adjacencies.} 
   The specification of $\bm{A}$ ensures $\mbox{corr}(\gamma_{i1}, \gamma_{i2})\approx 0.7$ between the two diseases.
   
   Figure~\ref{fig:boundary} shows the map for random effects for disease $1$ on the left and disease $2$ on the right. There are five different levels in total for both diseases with values $-2.67$, $-1.73$, $-0.98$, $0.42$ and $0.77$ ordered from the smallest to largest. As a result, we found $75$ ``true difference boundaries" delineating clusters with substantially different values for disease $1$ and $78$ ``true difference boundaries" for disease $2$. Moreover, there are $77$ cross-disease difference boundaries delineating random effects for disease $1$ from disease $2$ in neighboring regions, i.e. $\phi_{i1} \neq \phi_{j2}, i \sim j$, and $i < j$; there are $95$ cross-disease difference boundaries separating disease $2$ from disease $1$ in the neighboring regions, i.e. $\phi_{i2} \neq \phi_{j1}, i \sim j$, and $i < j$.
   
   \subsection{Model Comparison}
   Fixing the values of $\bm{\phi}$ generated as above, we simulated 
   {$50$} datasets for the outcome 
   . We analyzed the 
   {$50$} replicated datasets using \eqref{eq:post_dist} with vague priors specified in (\ref{eq: prior}) as $a_s = 2$, $b_s = 0.1$, $a_e = 2$, $b_e = 0.1$, $\sigma_\beta^2 = 1000$, $\alpha = 1$, $\nu = 2$ and $\bm{R} = \mbox{diag}(0.1, 0.1)$. The same set of priors were used for both MDAGAR and MCAR as they have the same number of parameters with similar interpretations. The joint multivariate settings were compared with corresponding independent-disease models for CAR and DAGAR respectively. For independent-disease models, spatial components are assumed to be independent between diseases. Hence $\displaystyle \bm{A} = \begin{pmatrix} a_{11} & 0 \\ 0 & a_{22} \end{pmatrix}$ and $\displaystyle \bm{\Sigma}_\gamma = \begin{pmatrix} a_{11}^2\bm{Q}^{-1}(\rho_1) & \bm{O} \\ \bm{O} & a_{22}^2\bm{Q}^{-1}(\rho_2)\end{pmatrix}$ is block diagonal
   . We refer to the independent-disease models by DAGAR$_{ind}$ and CAR$_{ind}$ according to whether $\bm{Q}(\rho_d)$ is specified by DAGAR and CAR, respectively. We used the same priors as for the joint models except for $\bm{A}$, which is now specified by $a_{dd}^2 \sim IG(a_v, b_v), a_v = 2, b_v = 0.1$ for $d=1,2$. 
   All models were executed
   in the \texttt{R} statistical computing environment and %
   inference was %
   obtained %
   from %
   $5000 \times 2\mbox{ (chains) } = 10000$ MCMC samples from (\ref{eq:post_dist}) for each model.
   
   We compared MDAGAR, MCAR, DAGAR$_{ind}$ and CAR$_{ind}$ using 
   a predictive loss criterion based on a balanced loss function for replicated data sets \citep{gelfand1998model}. For the latter, we drew replicates $y_{\text{rep},id}^{(\ell)} \sim N\left(\bm{x}_{id}^{\top}\bm{\beta}_d^{(\ell)} + \phi_{id}^{(\ell)}, 1/\tau_{d}^{(\ell)}\right)$ for each posterior sample $\ell = 1, \ldots, L$ and computed $D = G + P$, where $G = \sum_{d=1}^q \sum_{i=1}^n (y_{id} - \bar{y}_{\text{rep},id})^2$ and $P = \sum_{d=1}^q\sum_{i=1}^n\sigma_{\text{rep},id}^2$, $\displaystyle \bar{y}_{\text{rep},id} = \frac{1}{L}\sum_{\ell=1}^L y_{\text{rep},id}^{(\ell)}$, and $\sigma_{\text{rep},id}^2$ is the variance of $y_{\text{rep},id}^{(\ell)}$ for $\ell = 1, \ldots, L$. 
   $D$ 
   rewards goodness of fit and 
   penalizes model complexity. Figure~\ref{fig:model_fitting} plots values of 
   $D$ (\ref{fig: D}) over the 
   {$50$} data sets for the four models. The two joint models exhibit much better performance with lower 
   $D$ scores than the independent models. This, unsurprisingly, indicates the benefits of capturing dependence among diseases. MDAGAR and MCAR perform comparably, although CAR$_{ind}$ seems to be slightly preferred to DAGAR$_{ind}$. 
   
   
   We also computed the Kullback-Leibler Divergence, $D_{KL}\left(p(\bm{y}_{true})|| p(\bm{y})\right)$, between the true density $p(\bm{y}_{true})$ and the four models. Here, $p(\bm{y}_{true}) = N\left(\bm{y}_{true}\given \bm{X}\bm{\beta}_{true} + \bm{\phi}_{true}, diag(\bm{\sigma}_{true}) \otimes \bm{I}_n\right)$ and $p(\bm{y}) = N(\bm{y}\given \bm{X}\bm{\beta} + \bm{\phi}, diag(\bm{\sigma})\otimes \bm{I}_n)$ is the density from each candidate model, where $\mbox{diag}(\bm{\sigma})$ is a diagonal matrix with $1/\tau_d$ as $d$-th diagonal element, and $\bm{X}$ is a block diagonal design matrix with $\bm{X}_d = (\bm{x}_{1d}, \bm{x}_{2d}, \dots, \bm{x}_{nd})^\top$ as diagonal blocks. Since $D_{KL}\left(p(\bm{y}_{true})|| p(\bm{y})\right)$ is a function of the model parameters, we can compute its posterior distribution given each data set. We collect the posterior means from each dataset and plot them using a density-smoother in Figure~\ref{fig: KL} for the four models. These plots clearly show that the joint models, MCAR and MDAGAR, have smaller KL divergences from the true model than have CAR$_{ind}$ and DAGAR$_{ind}$. We also evaluated parameter estimates from the four models as discussed in Section~\ref{sup_sim}.
   
   Turning to boundary detection, we computed $P(\phi_{id} \neq \phi_{jd'} \given \bm{y})$ for $ d,d' = 1,2$ and for every pair of neighboring regions $(i, j)$. 
   Given these posterior probabilities, we obtained the corresponding boundary detection results (sensitivity and specificity) between and across diseases over our 
   {$50$} simulated datasets using our four models 
   {as well as the MBLV method}. Table~\ref{tab: rate} presents these results. Given the true number of difference boundaries, sensitivities and specificities were calculated by choosing difference boundaries as a fixed number of edges ranked in terms of the $T$ highest posterior probabilities. This was repeated for $T=60, 65, 70, 75, 80, 85$ for disease 1, disease 2 and disease 1 vs. 2, while $T = 70, 75, 80, 85, 90, 95$ were used for disease 2 vs. 1. Overall, the two joint models produce comparable detection rates and outperform the two independent-disease models 
   {as well as MBLV method} in terms of sensitivity and specificity under all scenarios. 
   {There is little difference between MDAGAR and MCAR, though MCAR performs slightly better when detecting boundaries between two diseases using larger $T$.} When $T$ is set close to the true number of difference boundaries for each disease, MDAGAR and MCAR are able to detect about $85\%$ of the true boundaries with specificity and sensitivity both around $85\%$ for disease 1, disease 2 and disease 1 vs. 2. When comparing diseases 2 vs. 1, MDAGAR and MCAR detect about $82\%$ of the true boundaries with specificity and sensitivity around $82\%$ when $T=85$. In most of these settings, the disease-independent models are more likely to produce false positives, recognizing the null case (i.e. $\phi_{id} = \phi_{jd'}$) as difference boundaries. 
   {The MBLV outperforms disease-independent models for boundary detection within each disease when $T$ is over $70$ edges since the underlying MCAR specification captures the dependence among diseases.}
   
\begin{figure}[H]
	\centering
	\includegraphics[width=120mm]{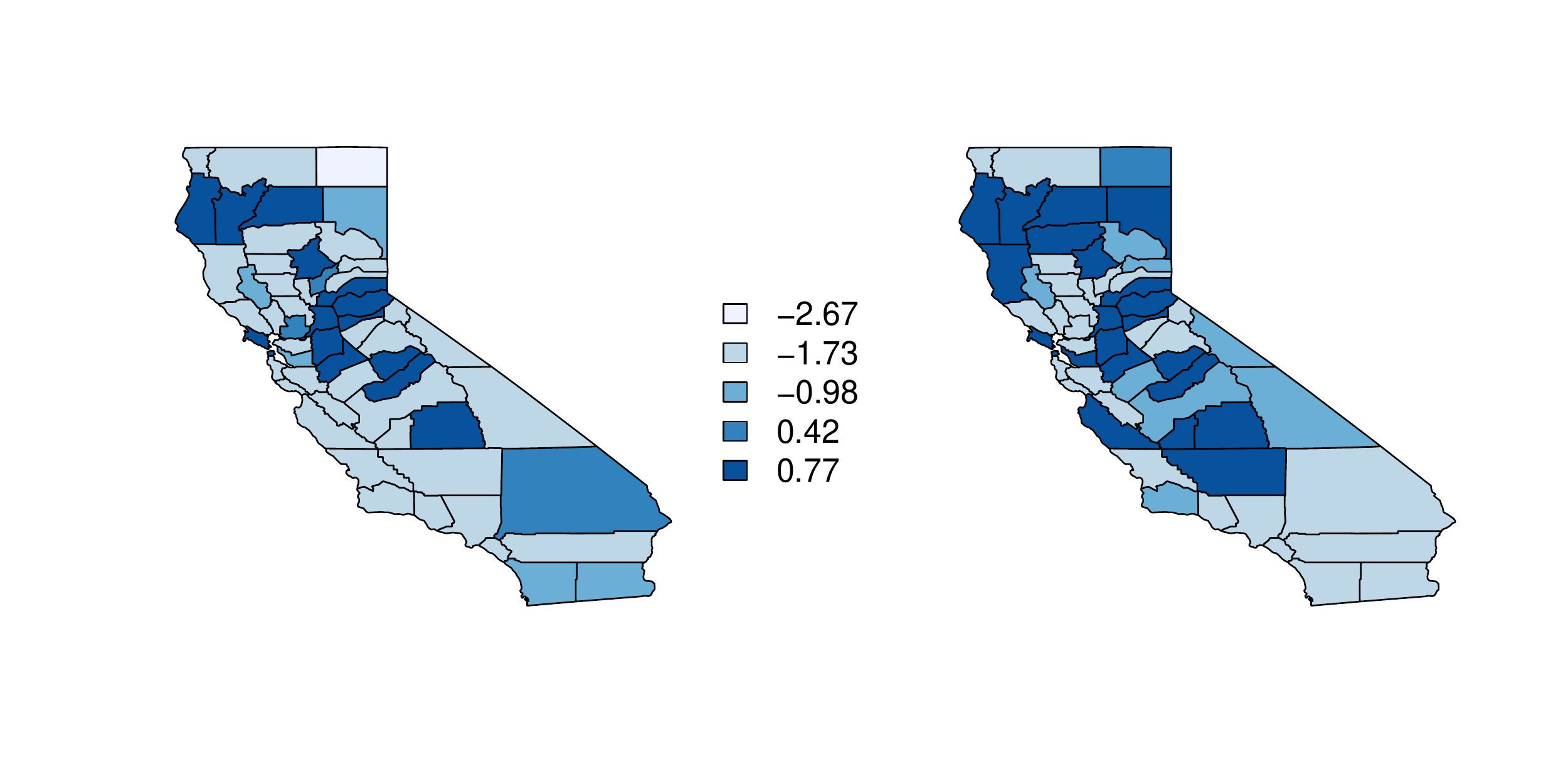}
	\caption{A map of the simulated data for random effects for disease $1$ (left) and disease $2$ (right) showing five different levels, each with its own value. There are $75$ boundary segments that separate regions for disease $1$ and $78$ difference boundaries for disease $2$.}\label{fig:boundary}
\end{figure}

\begin{figure}[H]
   	\hskip 0.2cm
   	\begin{subfigure}[t]{0.5\textwidth}
   		\centering
   		\includegraphics[width=75mm]{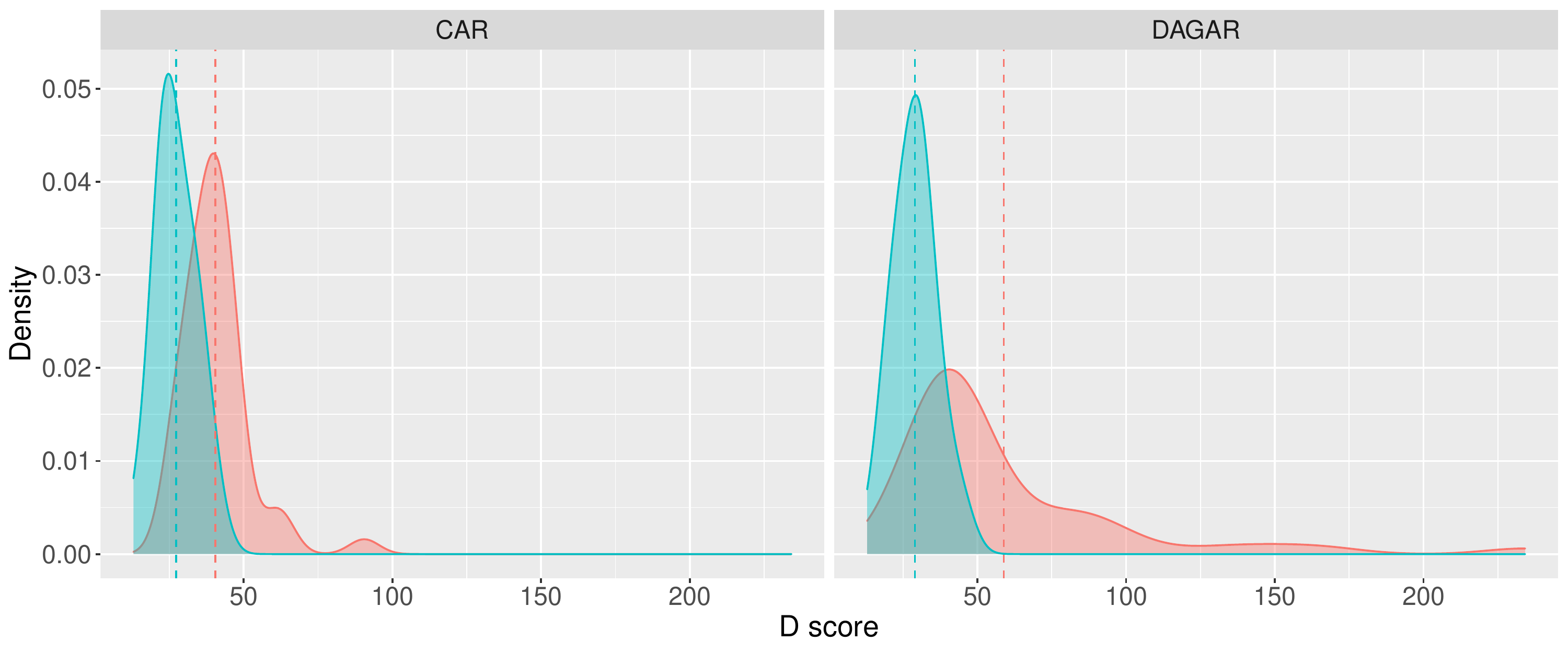}
   		\caption{D score}\label{fig: D}
   	\end{subfigure}
   	\begin{subfigure}[t]{0.5\textwidth}
   		\centering
   		\includegraphics[width=85mm]{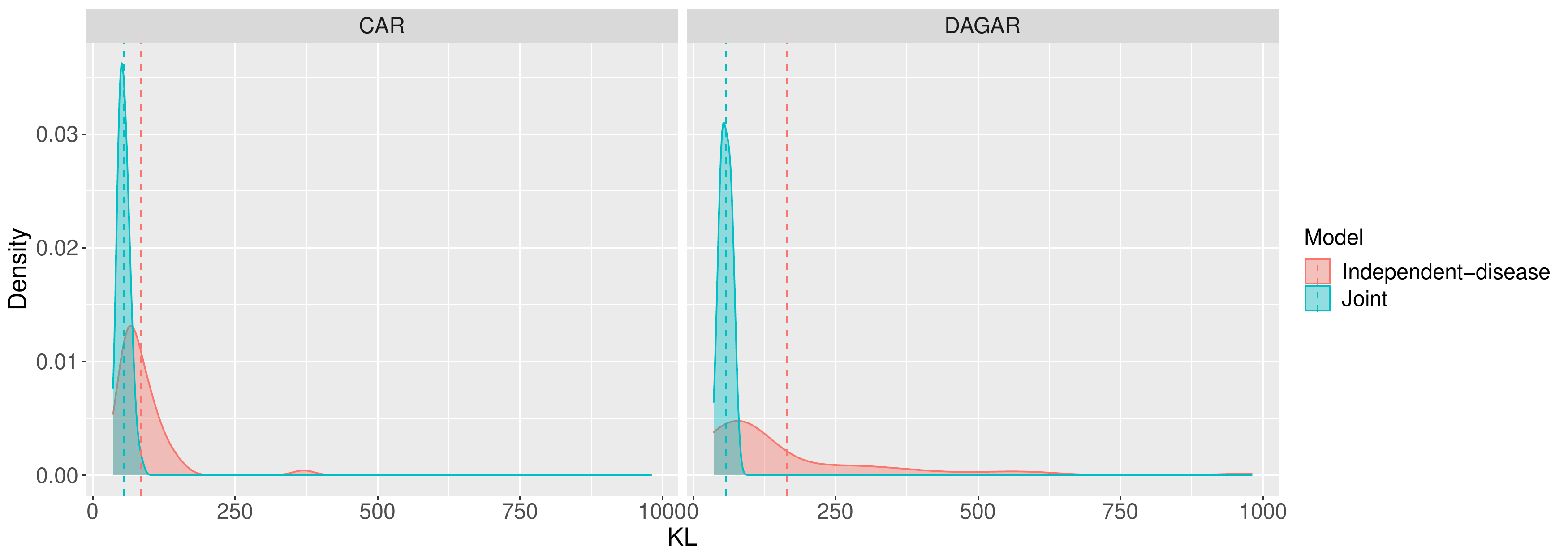}
   		\caption{KL}\label{fig: KL}
   	\end{subfigure}
   	\caption{Density plots for 
   	$D$ scores and mean $D_{KL}(p(\bm{y}_{true})|| p(\bm{y}))$ over 50 datasets as shown in (a)~and~(b) respectively, using two joint models, MCAR (blue plot in CAR panel) and MDAGAR (blue plot in DAGAR panel), and two independent-disease models, CAR$_{ind}$ (red plot in CAR panel) and DAGAR$_{ind}$ (red plot in DAGAR panel). The dotted vertical line shows the mean for each plot.}\label{fig:model_fitting}
   \end{figure}

   We next attend to detecting ``disease differences'' within the same county by computing $P(\phi_{i1} \neq \phi_{i2} \given \bm{y})$. This reflects difference in the random effects between two diseases in the same county. There are $20$ counties with true ``disease differences'' in Figure~\ref{fig:boundary}. Table~\ref{tab:disease_diff} shows sensitivity and specificity for detecting ``disease difference'' in the same county using the four models over 
   {$50$} datasets by choosing edges based upon the highest $T$ values of $P(\phi_{i1} \neq \phi_{i2} \given \bm{y})$. With $T=15, 20, 22, 25, 30$ we find, unsurprisingly, that MDAGAR and MCAR again excel over the two independent-disease models and MCAR-BLV in all scenarios with a resulting sensitivity and specificity of about $80\%$ when $T = 22$. 
   {Moreover, MDAGAR tends to have better specificity while MCAR tends to have higher sensitivity. 
   The MBLV performs poorly in detecting ``disease differences'' and boundaries across diseases (as shown in Table~\ref{tab: rate}), which is unsurprising given its limitations to account for propagation of uncertainties in boundary detection.}
   
\section{Analysis of SEER Dataset with Four Cancers}
   \label{data_analysis}
   \subsection{Data Example}
   \label{data}
\begin{figure}[H]
 	\centering
 	\includegraphics[width=120mm]{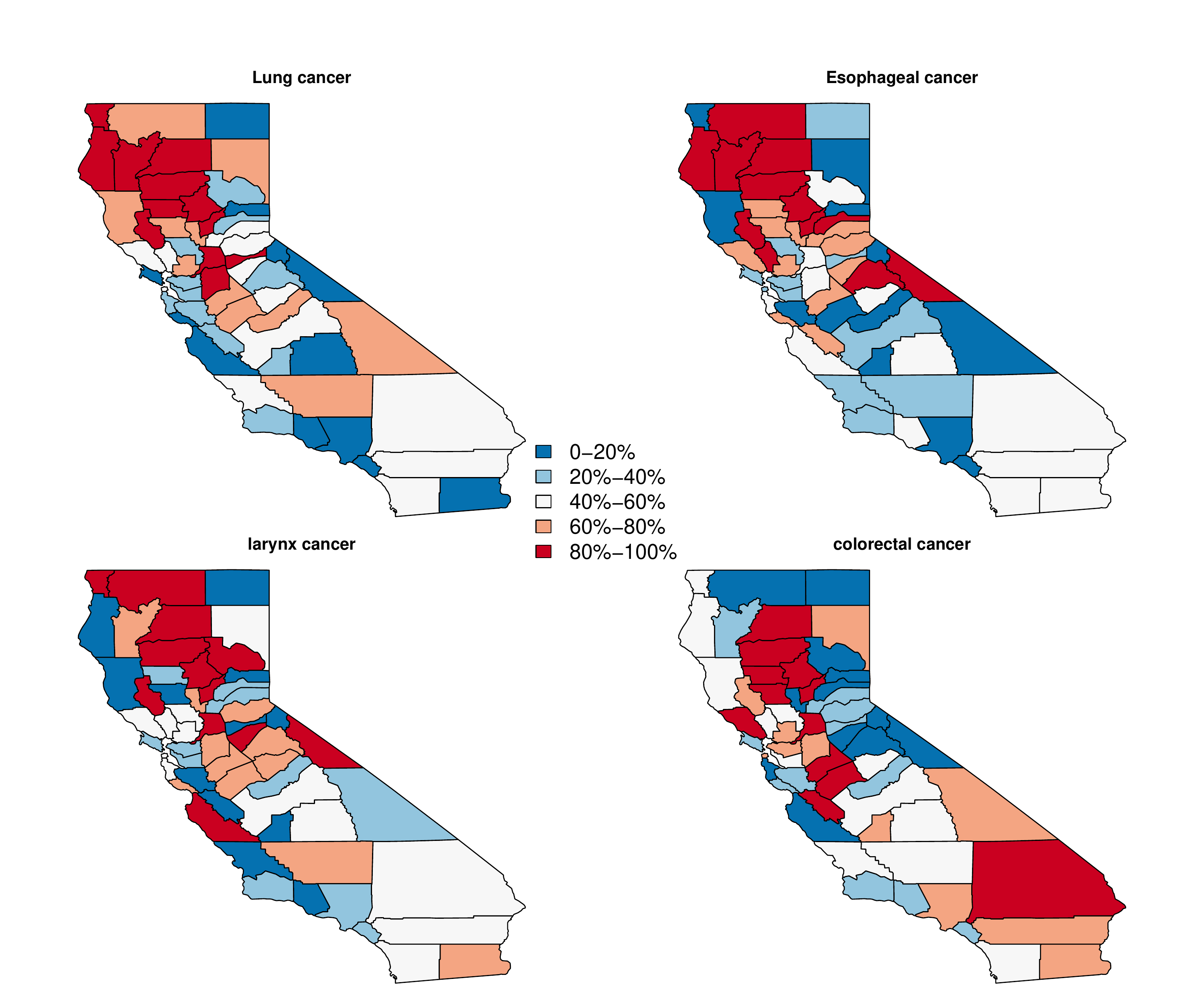}
 	\caption{Maps of age-sex adjusted standardized incidence ratios (SIR) for lung, esophageal, larynx and colorectal cancer in California, $2012 - 2016$.}\label{fig:raw_map}
 \end{figure}

   We consider an areal dataset recording the incidence of $4$ potentially interrelated cancers: lung, esophageal, larynx and colorectal. Lung and esophageal cancers have been found to share common risk factors \citep{agrawal2018risk} and metabolic mechanisms \citep{shi2004frequencies}. Lung cancer appears to be one of the most common second primary cancers in patients with colon cancer \citep{kurishima2018lung}. Additionally, patients with laryngeal cancer also have a high risk of developing second primary lung cancer \citep{akhtar2010second}. We extracted our data from the SEER$^*$Stat database using the SEER$^*$Stat statistical software \citep{seer}. The data consists of the observed counts of incidence ($Y_{id}$) for each cancer $d=1,2,3,4$ in each county $i=1,2,\ldots,58$ of California between $2012$ and $2016$. 
   {To calculate the expected number of cases $E_{id}$, we account for age-sex demographics in each county. We calculate the expected age-sex adjusted number of cases in county $i$ for cancer $j$ as $E_{id} = \sum_{k = 1}^m c_d^k N_i^k$, \citep[][]{jin2005generalized}, where $c_d^k = (\sum_{i = 1}^{58} Y_{id}^k)/ (\sum_{i = 1}^{58} N_i^k)$ is the age-sex specific incidence rate in age-sex group $k$ for cancer $d$ over all California counties, $Y_{id}^k$ is the counts of incidence in age-sex group $k$ of county $i$ for cancer $d$ and $N_i^k$ is the population in age-sex group $k$ of county $i$. The age groups are defined using 5 year increments until 85+ years (less than 1 year, 1 - 4 years, 5 - 9 years, 10 - 14 years, \dots, 80 - 84 years, 85+ years) and there are $m = 19 * 2 = 38$ age-sex groups. The age-sex adjusted standardized incidence ratios (SIR$_{id} = Y_{id}/E_{id}$) is plotted on a county map of California map in Figure~\ref{fig:raw_map}.} Cutoffs for the different levels of SIRs are quintiles for each cancer.
   
\begin{figure}[H]
 	\centering
 	\includegraphics[width=100mm]{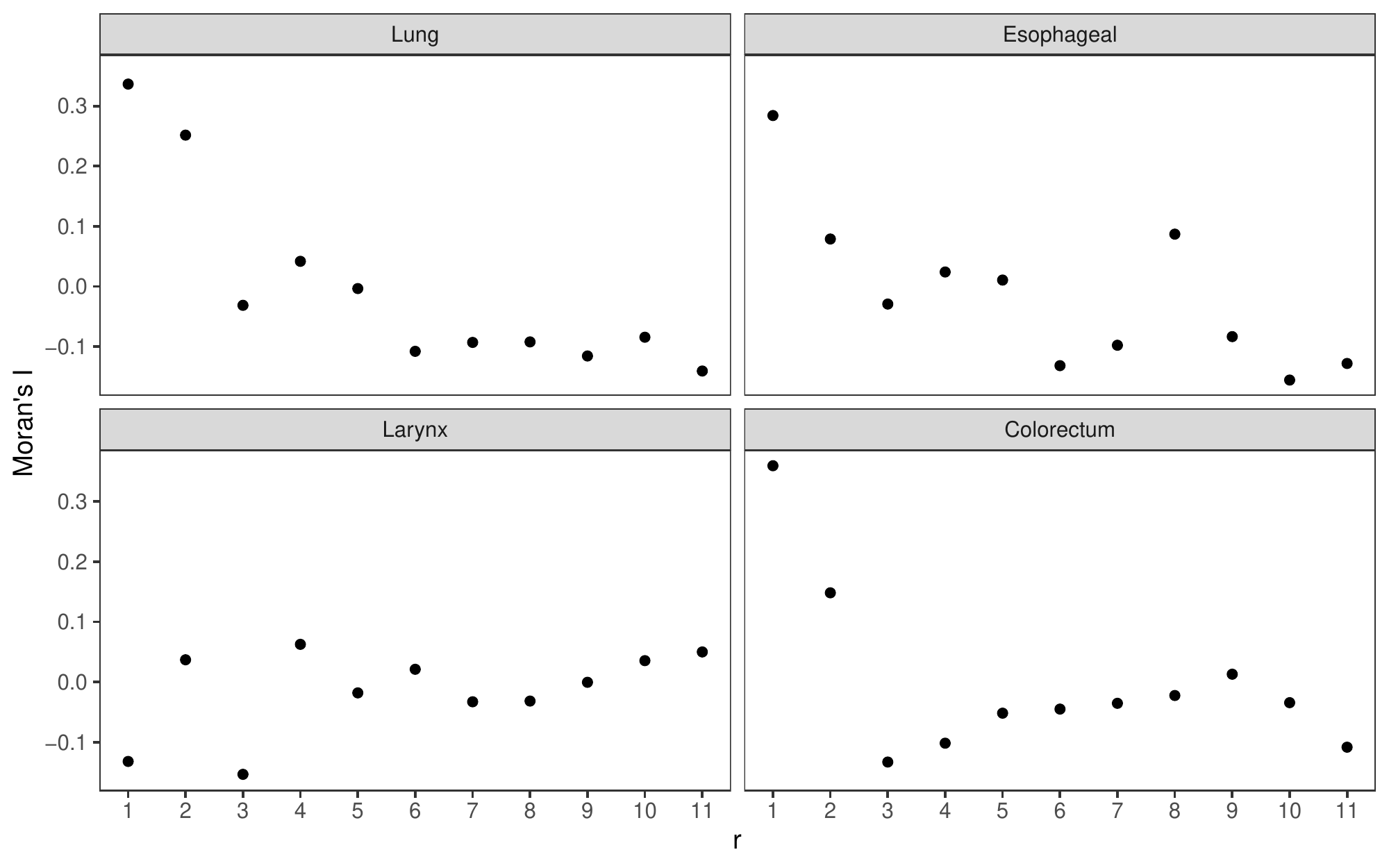}
 	\caption{Moran's I of $r$th order neighbors for lung, esophageal, larynx and colorectal cancer.}\label{fig: MoranI}
 \end{figure}

   As an exploratory tool to assess associations among the cancers, we calculated Pearson's correlation for each pair of cancers by regarding SIRs in different counties as independent samples and found that 
   {the incidence of lung cancer is significantly associated with esophageal, larynx and colorectal cancer with correlations of 0.58, 0.40 and 0.5 respectively. Meanwhile, the correlation between esophageal and larynx cancer is 0.42.} Next, to explore the spatial association for each cancer, we calculated Moran's I based upon the $r$th order neighbors for each cancer and plotted the areal correlogram \citep{banerjee2014hierarchical}. Defining distance intervals as $(0, d_1], (d_1, d_2], (d_2, d_3], \dots$, the $r$th order neighbors refer to units with distance in $(d_{r-1}, d_r]$, i.e. within distance $d_r$ but separated by more than $d_{r-1}$. The distance is the Euclidean distance from an Albers map projection of California. Figure~\ref{fig: MoranI} reveals that spatial associations in lung, esophageal and colorectal cancers clearly diminish with increasing $r$, although the pattern is less pronounced for larynx.
   
\begin{figure}[H]
 	\centering
 	\includegraphics[width=100mm]{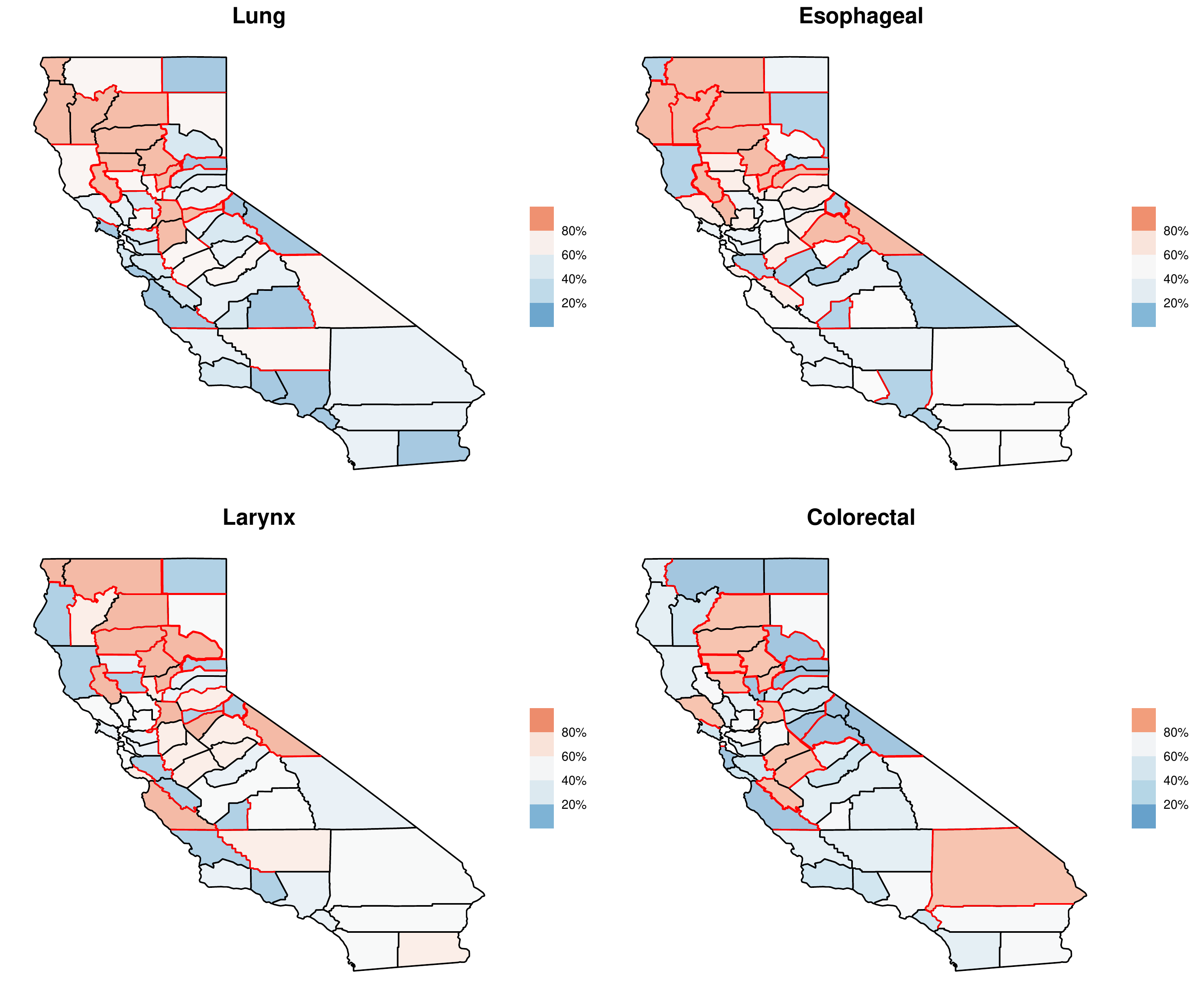}
 	\caption{Boundaries (in red) selected as the first $70$ pairs with largest differences for lung, esophageal, larynx and colorectal cancer.}\label{fig: raw_diff}
 \end{figure}

   For insights into difference boundaries for each cancer, we calculated the difference in SIR between each pair of neighboring counties ($139$ pairs in total), i.e., $|SIR_{id} - SIR_{jd}|, i \sim j$. By ranking the differences from largest to smallest, we selected the first $70$ pairs (half of the total pairs) with the largest differences as the difference boundaries for each cancer as shown in Figure~\ref{fig: raw_diff}. The four cancers exhibit similar patterns in boundary detection that more boundaries are detected in the north and the borders of California. Counties along the central corridor of California, ranging from central to south, tend to be in the same cluster. 
   
   \subsection{Data Analysis} \label{analysis}
   We analyzed the dataset mentioned in Section~\ref{data} using a Poisson spatial regression model, i.e. $Y_{id} \overset{\text{ind}}{\sim} Poisson\left(E_{id} \exp\left({\bm{x}_{id}^\top\bm{\beta}_d + \phi_{id}}\right)\right)$ for $i = 1, \ldots, 58$ and $d=1,\ldots, 4$. Applying prior specification as in the simulation study, we implemented MARDP (recall Section~\ref{implementation}) using MDAGAR and MCAR. Posterior inference is based upon $10000$ MCMC samples after $20000$ iterations of burn-in for diagosing convergence.
   
\begin{figure}[H]
 	\centering
 	\hskip -0.5cm	\begin{subfigure}[t]{0.5\textwidth}
 		\centering
 		\includegraphics[width=75mm]{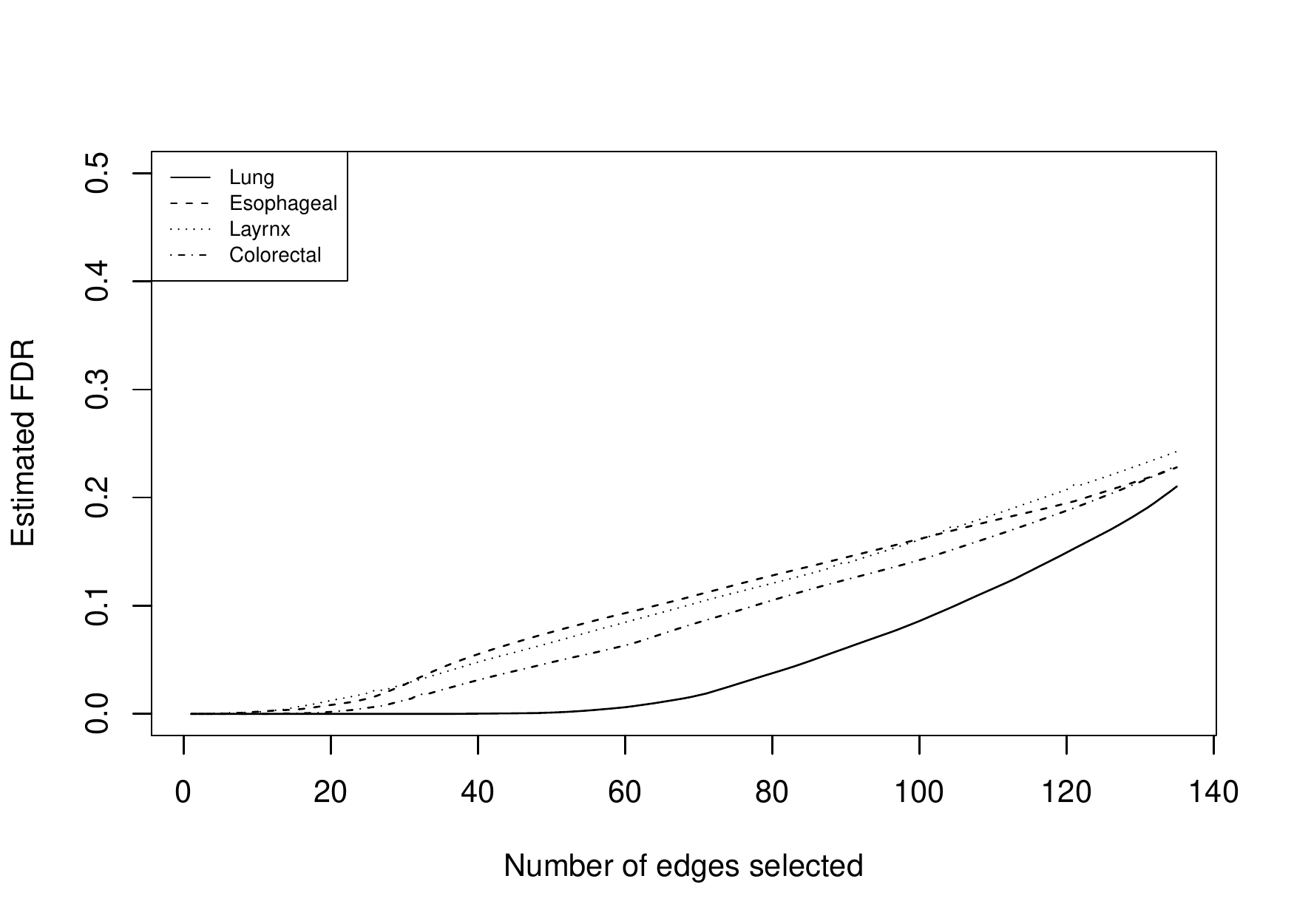}
 		\caption{MDAGAR}\label{fig: fdr_d}
 	\end{subfigure} 
 	\begin{subfigure}[t]{0.5\textwidth}
 		\centering
 		\includegraphics[width=75mm]{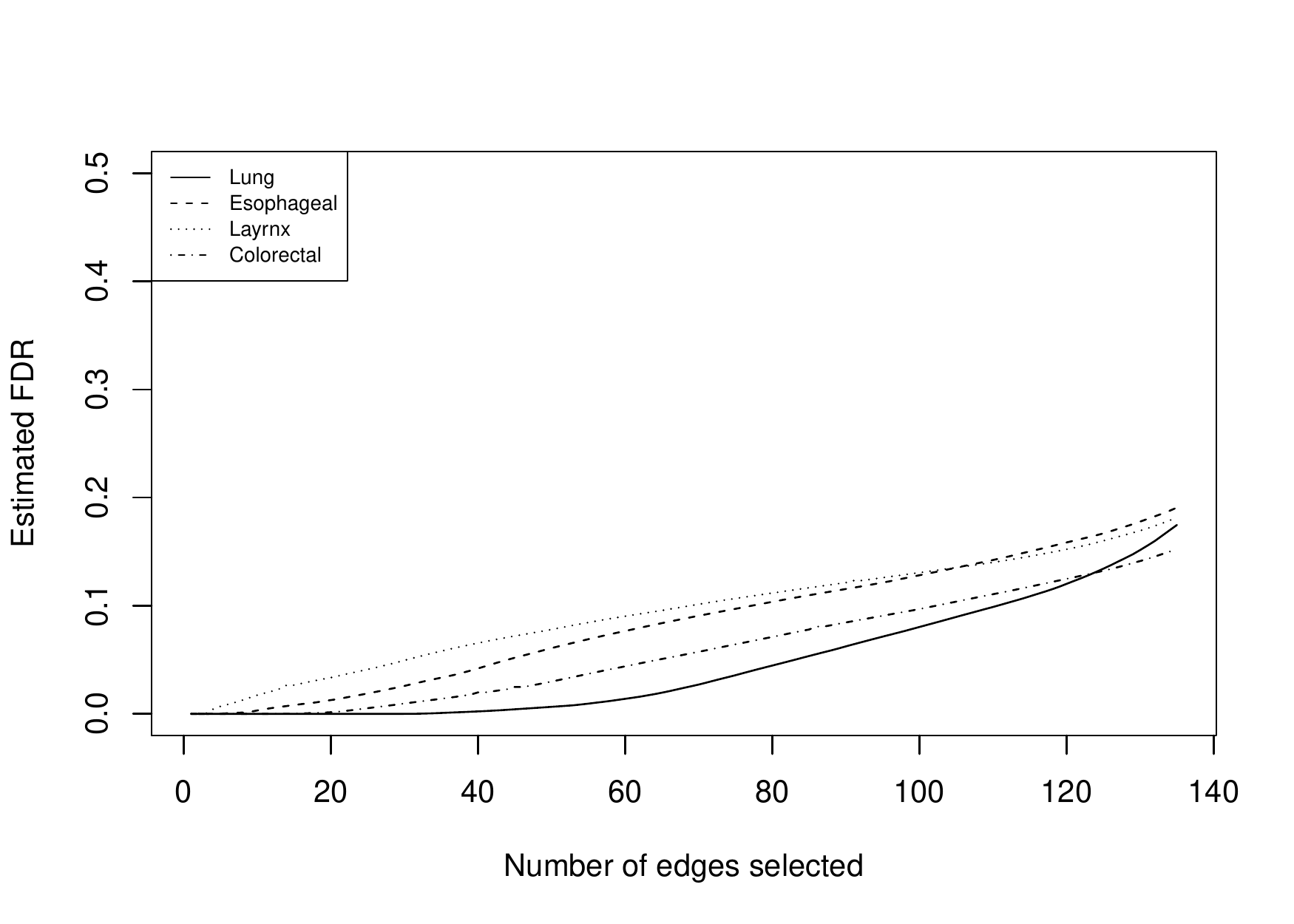}
 		\caption{MCAR}\label{fig: fdr_c}
 	\end{subfigure} 
 	\caption{Estimated FDR curves plotted against the number of selected difference boundaries for four cancers using MDAGAR and MCAR.}\label{fig:FDR}
 \end{figure}

\begin{figure}[H]
 	\centering
 	\begin{subfigure}{0.5\linewidth}
 		\centering
 		\includegraphics[width=100mm]{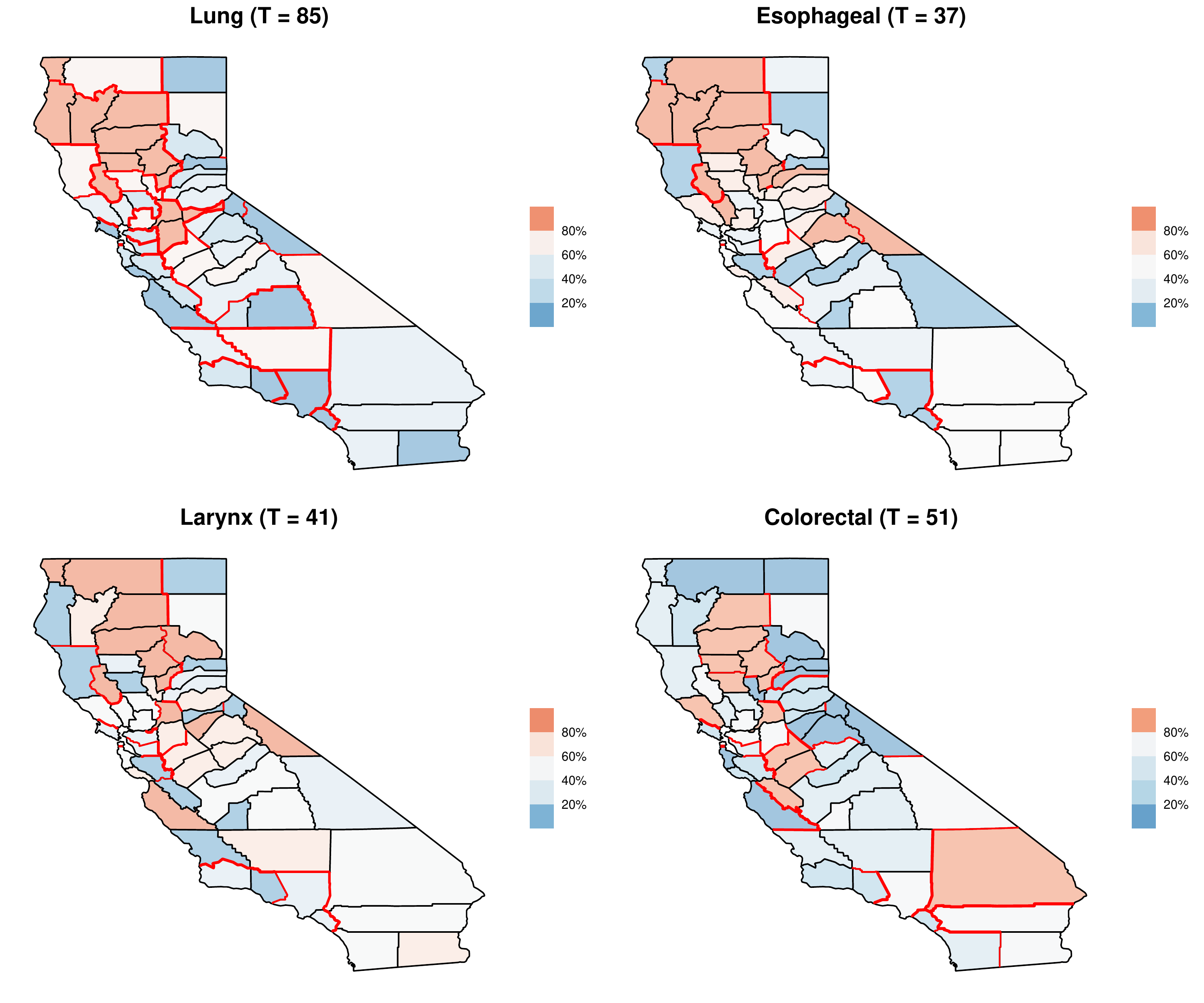}
 		\caption{MDAGAR}\label{fig: MDAGAR}
 	\end{subfigure}
 	\vfill
 	\vskip 1cm
 	\begin{subfigure}{0.5\linewidth}
 		\centering
 		\includegraphics[width=100mm]{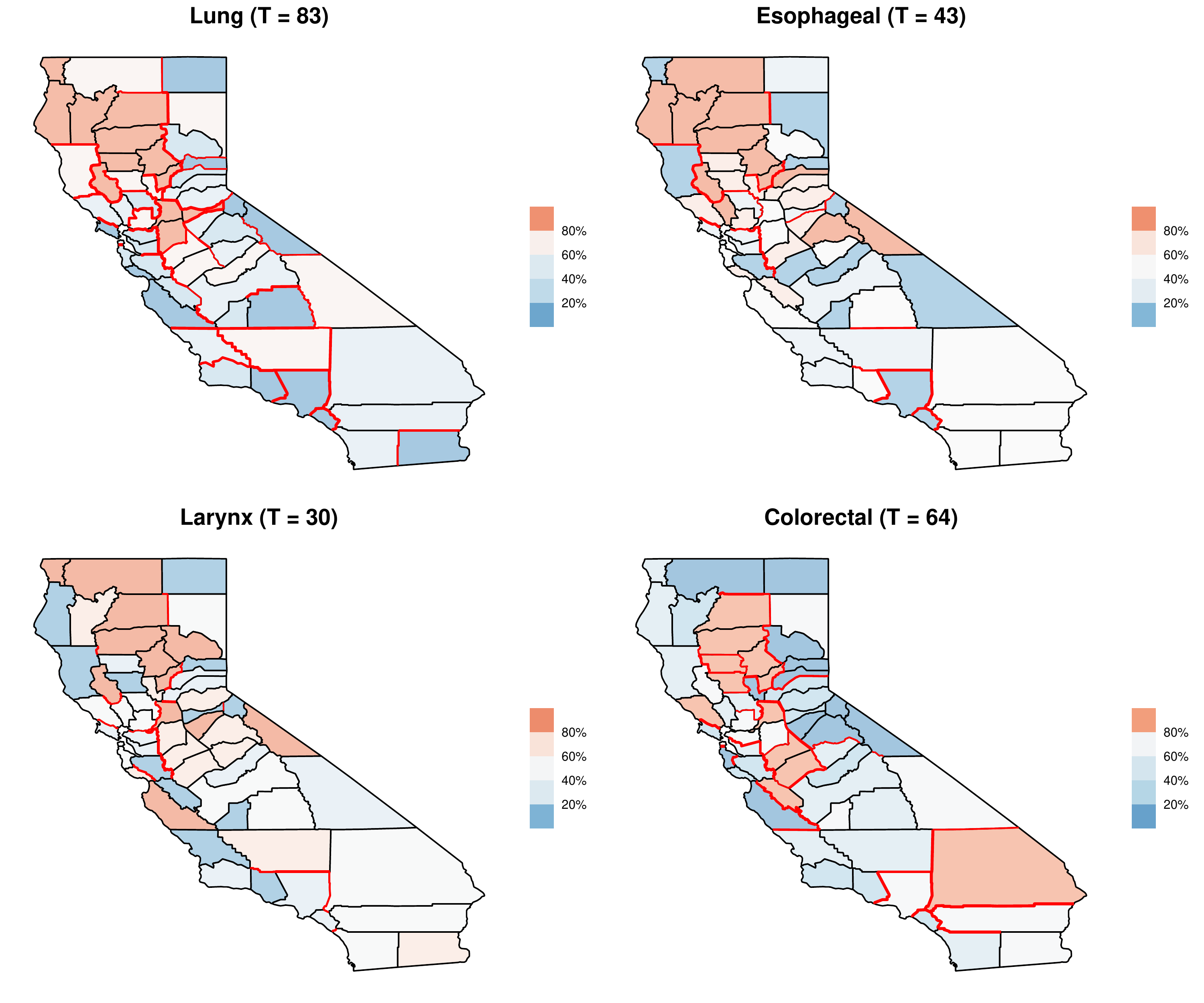}
 		\caption{MCAR}\label{fig: MCAR}
 	\end{subfigure}
 	\caption{Difference boundaries (highlighted in red) detected by (a) MDAGAR and (b) MCAR in SIR map for four cancers individually when $\delta = 0.05$. The values in brackets are the number of difference boundaries detected.}\label{fig: diff_bound}
 \end{figure}

   Without accounting for covariates, we detected difference boundaries for SIR of each cancer and across cancers. First, regarding boundary detection for each cancer, 
   we set up a threshold to control for FDR as in \eqref{eq: threshold}. Figure~\ref{fig:FDR} plots the change of estimated FDR with different numbers of edges selected as difference boundaries for the four cancers individually using MDAGAR (\ref{fig: fdr_d}) and MCAR (\ref{fig: fdr_c}). In general, MDAGAR and MCAR render similar trends in FDR curves, which are close to each other for esophageal, colorectal and larynx cancers while lung cancer exhibits much smaller values. 
   {The FDR increases slightly faster for esophageal cancer with MDAGAR and for larynx cancer with MCAR. We detect more boundaries for lung and fewer boundaries for esophageal and larynx cancer using the same threshold.} Setting 
   {$\delta = 0.05$} in \eqref{eq: threshold}, Figure~\ref{fig: diff_bound} shows difference boundaries (highlighted in red) detected by MDAGAR and MCAR in SIR maps for the four cancers. 
   {Wider lines indicate more prominent boundaries with higher probabilities of detection.}
   Maps from MDAGAR and MCAR are consistent with each other 
   {with similar boundary patterns} and the number of difference boundaries detected by the two models are also similar for each cancer, 
   {albeit with more boundaries detected for larynx ($41$ edges with posterior probabilities above the threshold $t^\star$ in \eqref{eq: threshold}) and fewer boundaries detected for colorectal ($51$ edges) under MDAGAR.} For lung cancer around $85$ boundaries are detected, which is considerably higher than the other three cancers. 
   
   Table~\ref{tab:bound_name} provides an exhaustive list of the cancer boundaries detected by MDAGAR in Figure~\ref{fig: diff_bound}. This ``lookup table'' contains the names of adjacent counties ranked in decreasing order of $P(\phi_{id} \neq \phi_{jd} \given \bm{y})$ for the four cancers, offering a detailed reference for health administrators to identify substantial spatial health barriers. Around $70\% - 90\%$ of the boundaries listed here are also detected by MCAR. For each cancer, we see some clusters and islands (regions fully encompassed by difference boundaries with all neighbors within California) in the map. 
   {For example, the northern counties of Shasta, Tehama, Glenn, Butte, Humboldt and Trinity appear to form a cluster with larger effects for all cancers. Similarly, the central and southern counties of Merced, Mariposa, Madera, Fresno, Kings, Tulare and Inyo appear in the same cluster with moderate to lower effects for esophageal, larynx and colorectal cancers. Orange is different from all its neighbors for the effects of lung and colorectal cancers, while San Bernardino differs from all its neighbors with a larger effect for colorectal cancer. Meanwhile, Santa Barbara and Ventura form a cluster to form a difference boundary island for lung and esophageal cancers.} A map of California with names and geographic boundaries for each county is shown in Figure~\ref{fig: county_map} for reference.
   
\begin{figure}[H]
 	\centering
 	\includegraphics[width=140mm]{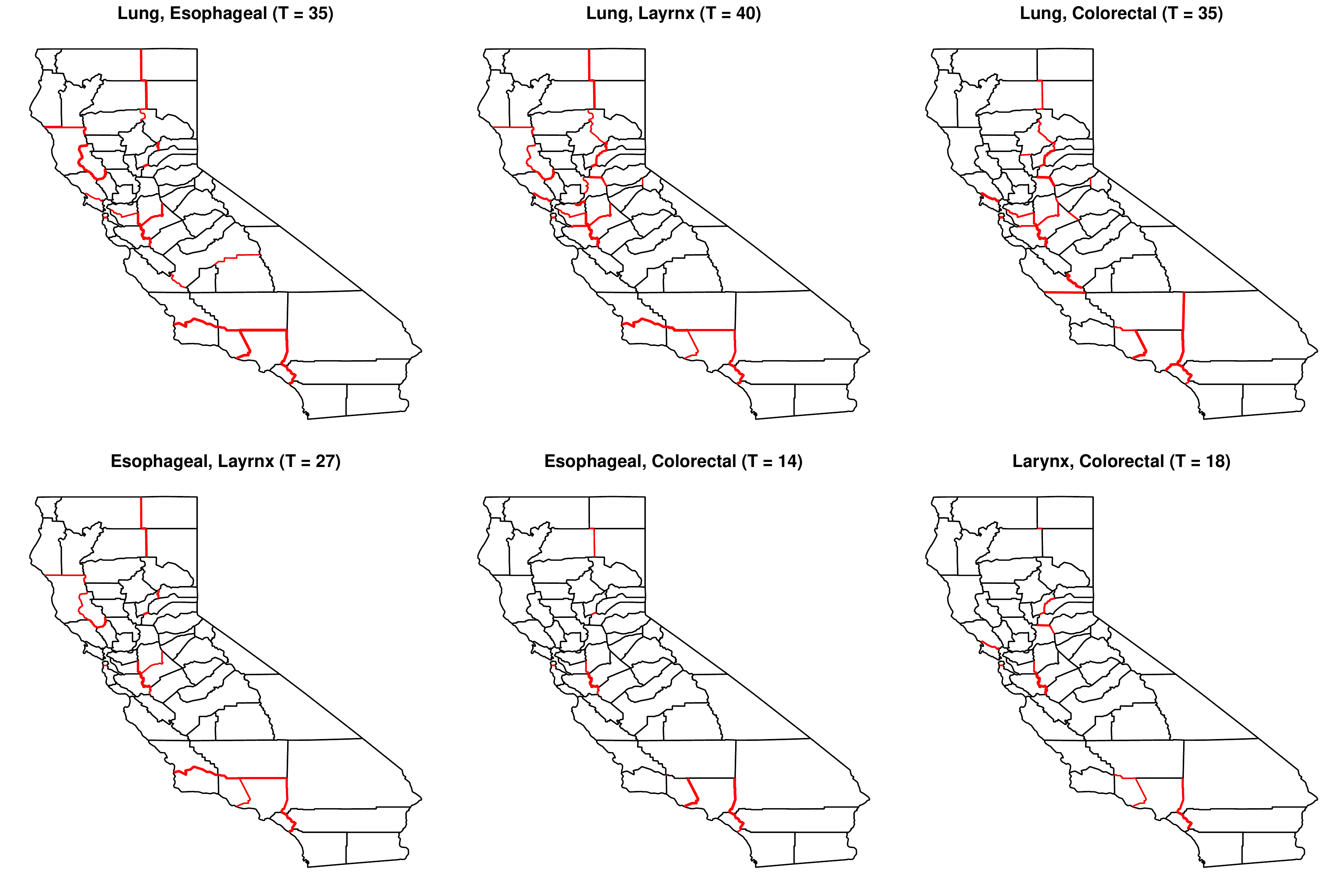}
 	\caption{Shared difference boundaries (highlighted in red) detected by MDAGAR for each pair of cancers in SIR map when $\delta = 0.05$. The values in brackets are the number of difference boundaries detected.}\label{fig: share}
 \end{figure}

\begin{figure}[H]
 	\centering
 	\includegraphics[width=140mm]{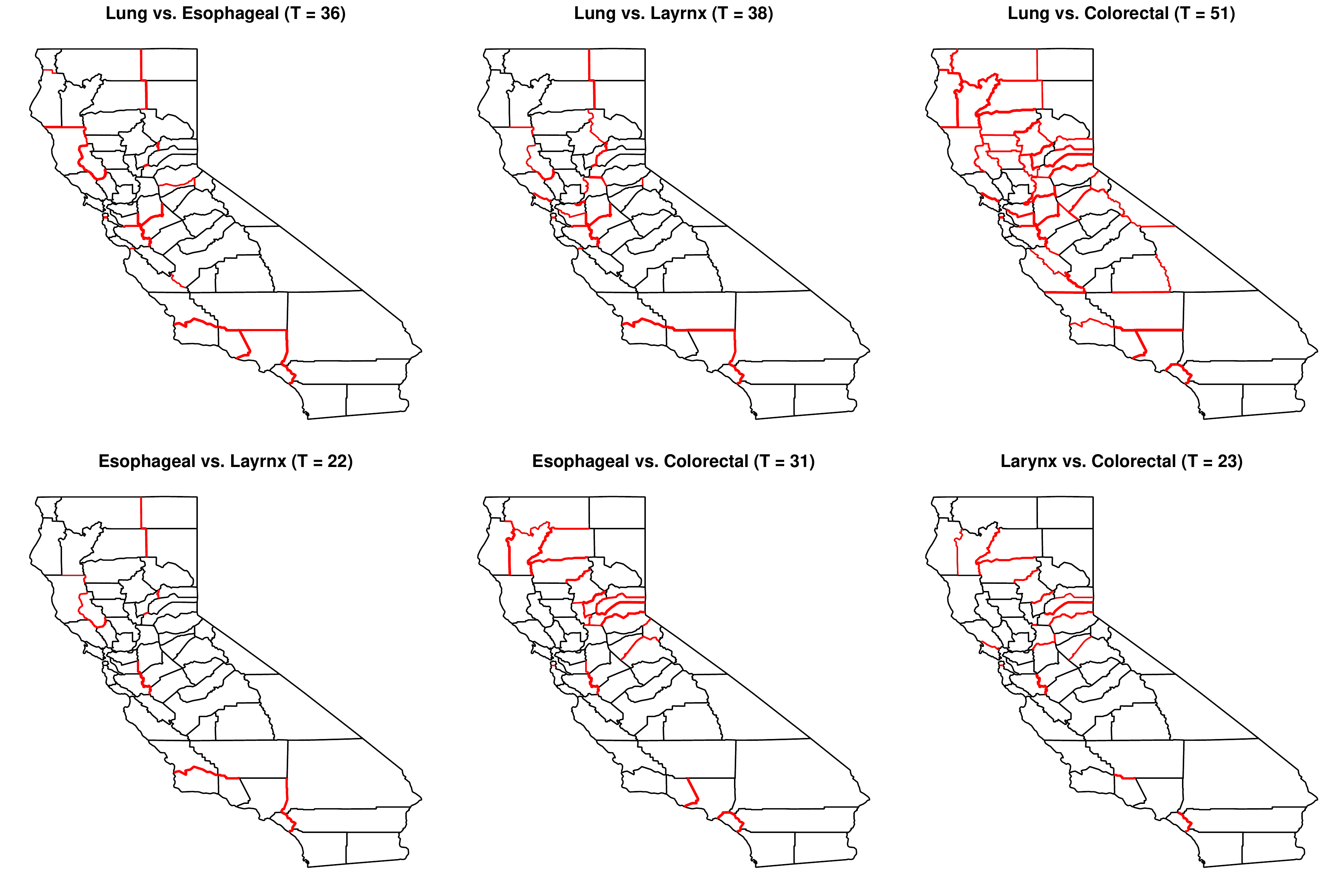}
 	\caption{Mutual cross-cancer difference boundaries (highlighted in red) detected by MDAGAR for each pair of cancers in SIR map when $\delta = 0.05$. The values in brackets are the number of difference boundaries detected.}\label{fig: cross}
 \end{figure}

   For difference boundaries between cancers, we considered the shared difference boundaries and cross-cancer boundaries. Here, we only show results from MDAGAR. The shared difference boundaries are defined as common boundaries detected for different cancers. Figure~\ref{fig: share} exhibits the shared boundaries for each pair of cancers, i.e. $P(\phi_{id} \neq \phi_{jd},  \phi_{id'} \neq \phi_{jd'}| \bm{y}), d \neq d'$. Consistent with results for individual cancers in Figure~\ref{fig: diff_bound}, 
   {Orange is the island with shared difference boundaries within California for [lung, esophageal]. Santa Barbara and Ventura together form an island for [lung, esophageal], [lung, larynx] and [esophageal, larynx]. Meanwhile, lung, esophageal and larynx cancers share difference boundaries between Lake and its three neighboring counties: Mendocino, Sonoma and Napa.} For cross-cancer difference boundaries, we define a mutual cross-cancer boundary from $P(\phi_{id} \neq \phi_{jd'}, \phi_{id'} \neq \phi_{jd} | \bm{y}), i \sim j, i < j$, which separates effects for different cancers mutually in neighboring counties (see Figure~\ref{fig: cross}). In conjunction with Figure~\ref{fig: diff_bound}, 
   {we observe that the shared difference boundaries for [lung, esophageal], [lung, larynx] and [esophageal, larynx] also tend to be mutual cross-cancer difference boundaries for the same pair. This indicates high correlation between the SIR's for lung, esophageal and larynx cancers. Compared with shared difference boundaries, mutual cross-cancer difference boundaries are detected between colorectal and the other three cancers indicating a different spatial pattern for colorectal cancer.}
   
   We also compare the two joint models with the two independent models. Table~\ref{tab:D_model} presents the predictive loss criterion $D$ score for the models. For Poisson regression, replicates for each data point are replaced by $y_{\text{rep},id}^{(\ell)} = Y_{\text{rep},id}^{(\ell)} / E_{id}$, where $Y_{\text{rep},id}^{(\ell)} \sim Poisson\left(E_{id}\exp\left({\bm{x}_{id}^\top\bm{\beta}_d^{(\ell)} + \phi_{id}^{(\ell)} }\right)\right)$. The $D$ scores are calculated for each cancer and added up for the four cancers to produce $D_{\text{sum}}$. 
   {It reveals that all four models perform competitively in terms of data fitting.} DAGAR$_{ind}$ and CAR$_{ind}$ detect fewer difference boundaries for each cancer under the same FDR threshold compared with MDAGAR and MCAR. 
   {When $\delta = 0.1$, DAGAR$_{ind}$ and CAR$_{ind}$ produce similar patterns with a similar number of boundaries as detected by MDAGAR and MCAR with $\delta = 0.05$ for lung and colorectal cancer (see Figure~\ref{fig: diff_bound}); fewer boundaries are detected for esophageal and larynx cancers.} 
   Detecting the shared boundaries between the three cancers pairwise using DAGAR$_{ind}$ and CAR$_{ind}$ under the same setting ($\delta = 0.1$) reveals fewer shared boundaries. 

We explore the impact of risk factors in boundary detection by including a potential common risk factor for cancers, adult smoking rates (smoking$_{id}$), for 2014--2016 obtained from the California Tobacco Facts and Figures 2018 database \citep{california2018california}, and 
{ 
percentage of unemployed residents (unemployed$_{id}$ in a county). This county attribute is common for different cancers and extracted from the SEER$^*$Stat database \citep{seer} for the same period, 2012--2016. Maps of these two covariates are shown in Figure~\ref{fig: covariates} using quintiles as cutoffs.} 

{Including covariates can result in detection of larger or smaller numbers of difference boundaries. For example, if including a covariate increases the difference between the values of the residual spatial effects between two neighboring counties, then including the covariate will tend to evince a difference boundary between those two neighboring counties. The reverse effect, i.e., including a covariate causes a difference boundary to disappear, will occur if it reduces the difference in values of residual spatial effects between neighboring counties. While including covariates will always absorb some spatial effects, they could increase or decrease the number of difference boundaries depending upon how they impact the \emph{difference} in rates across neighboring counties.} 

\begin{figure}[H]
	\centering
	\begin{subfigure}{0.5\linewidth}
		\centering
		\includegraphics[width=72mm]{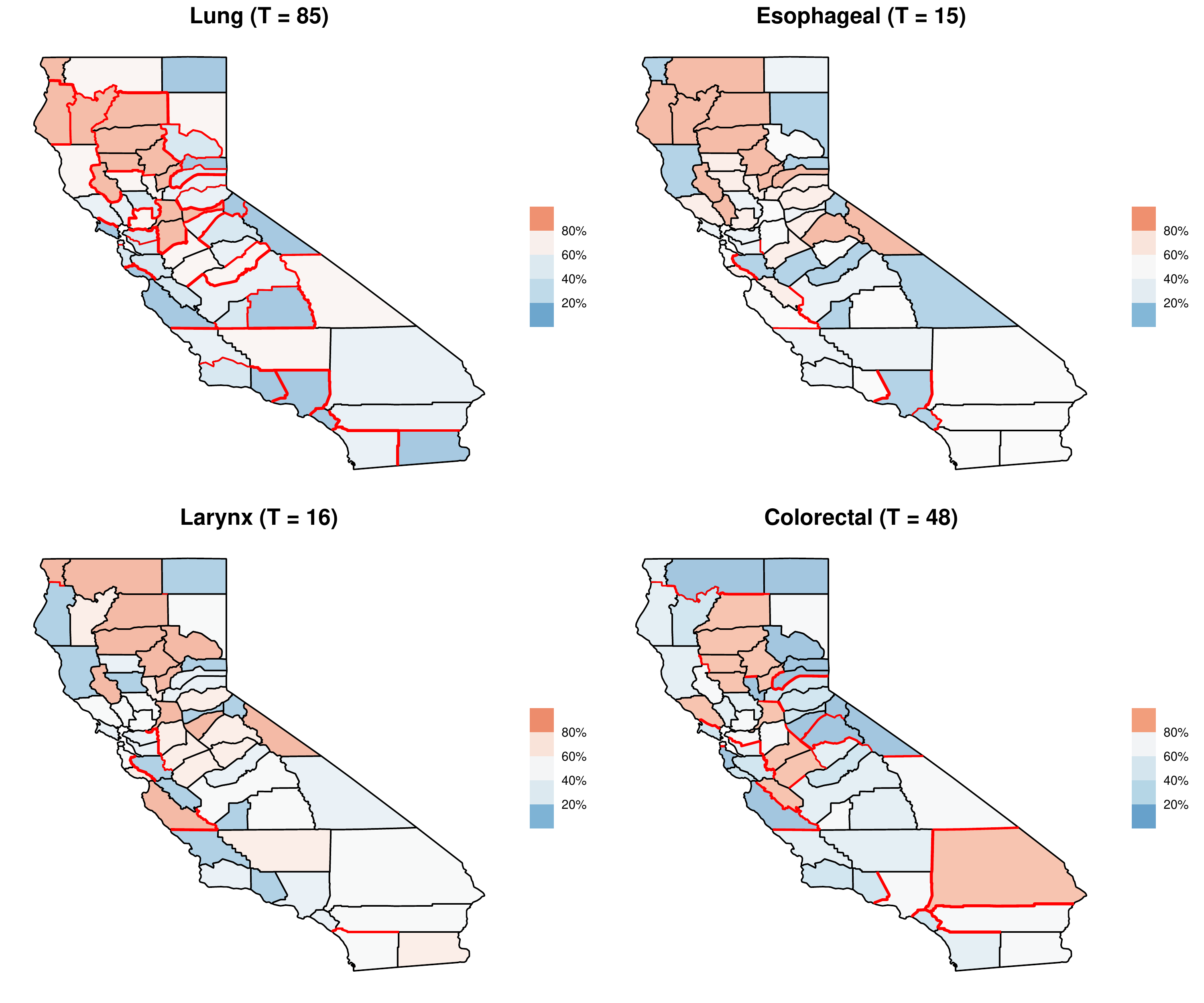}
		\caption{Smoking}\label{fig: fit_smoking}
	\end{subfigure}
	\hfill
	\hskip -6.5cm\begin{subfigure}{0.5\linewidth}
		\centering
		\includegraphics[width=72mm]{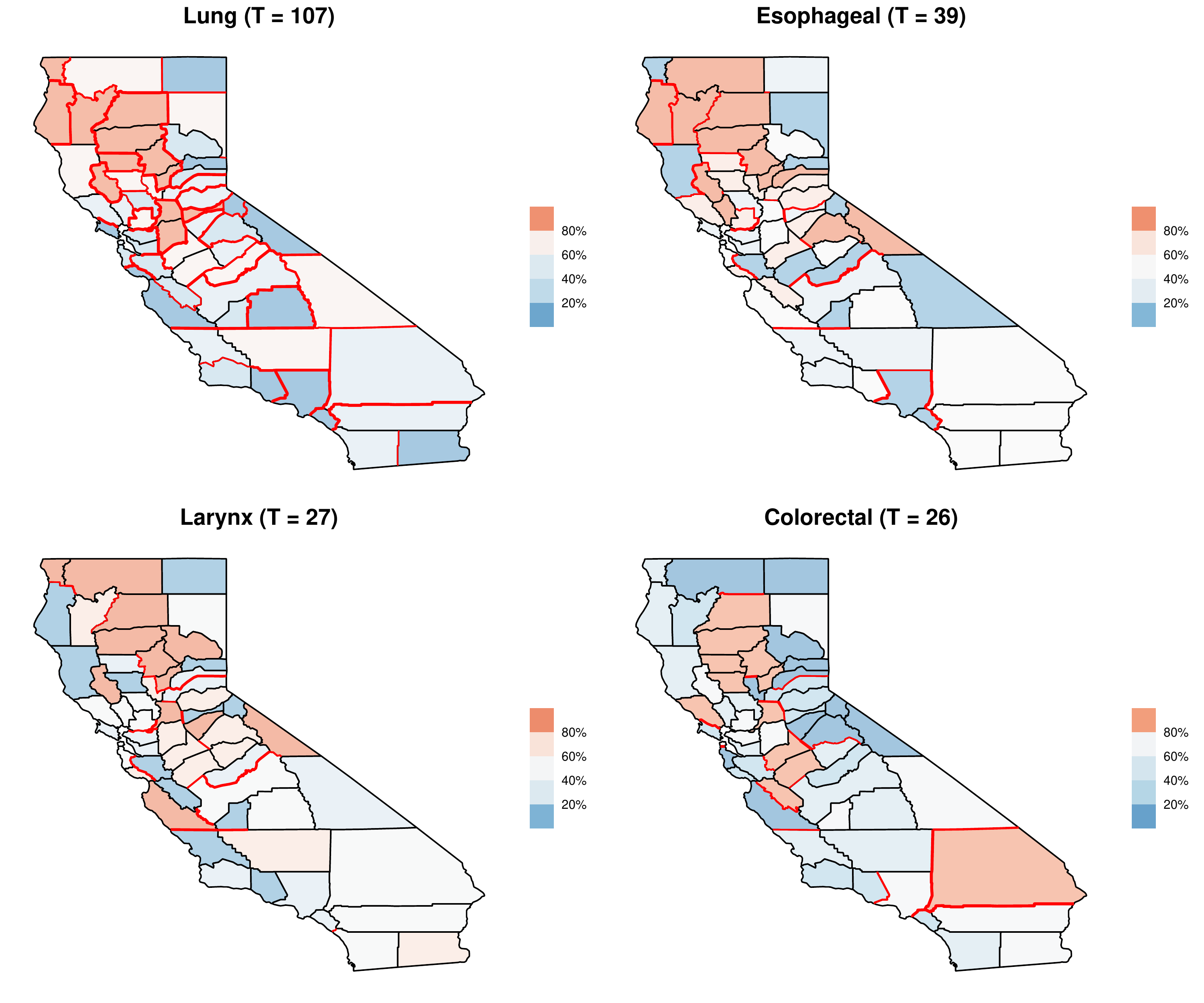}
		\caption{Smoking, Unemployed}\label{fig: fit_emp}
	\end{subfigure}
	\caption{Difference boundaries (highlighted in red) detected by MDAGAR after accounting for (a) smoking, (b) smoking and unemployed for four cancers individually when $\delta = 0.05$. The values in brackets are the number of difference boundaries detected.}\label{fig: diff_bound_cov}
\end{figure}

\begin{figure}[H]
	\centering
	\includegraphics[width=120mm]{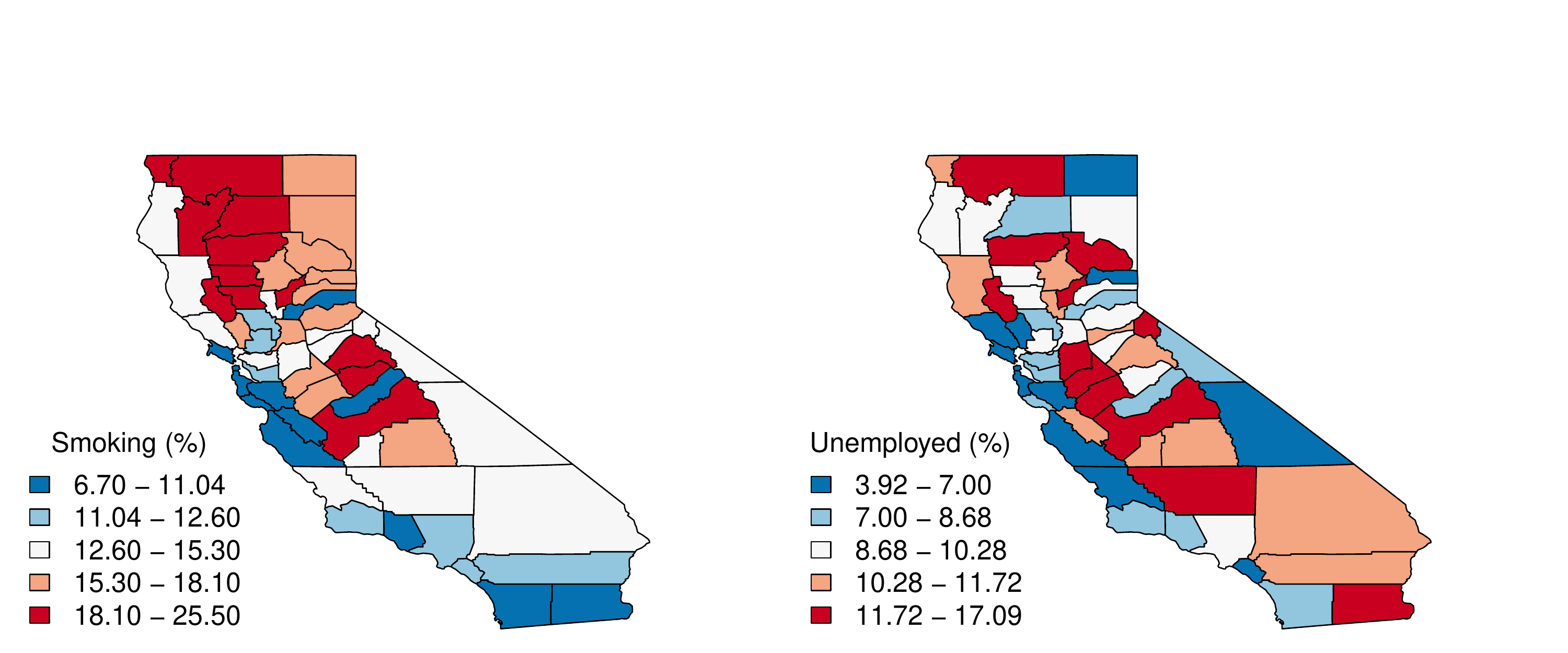}
	\caption{Maps of county-level covariates: adult cigarette smoking rates (left) and unemployed residents (right).}\label{fig: covariates}
\end{figure}

{Adding the two covariates sequentially, Figure~\ref{fig: diff_bound_cov} shows difference boundaries for all four cancers detected by MARDP with MDAGAR after accounting for only ``smoking'' in Figure~\ref{fig: fit_smoking}; and accounting for both ``smoking'' and ``unemployed'' in Figure~\ref{fig: fit_emp} when $\delta = 0.05$. Table~\ref{tab:coef} presents posterior means (95\% credible intervals) for regression coefficients and autocorrelation parameters estimated without any of the covariates (only an intercept), and sequentially adding the covariates (``smoking'' and ``unemployed''). Unsurprisingly, regression slopes for the percentage of smokers are significantly positive for all cancers when accounting for ``smoking'' only, while this effect is mitigated for colorectal cancer after introducing ``unemployed''. The percentage of unemployed residents also has a positive association with incidence rates for lung, larynx and colorectal cancer after controlling for ``smoking''. We also find that the spatial autocorrelation $\rho_d$ corresponding to the latent factor $\bm{f}_d$ varies considerably by cancer after accounting for two covariates. Larger estimates of $\rho_d$ imply smoother maps and, consequently, fewer difference boundaries.}

Compared to difference boundaries for SIR in Figure~\ref{fig: MDAGAR} without any covariates, we tend to find lower numbers of boundaries detected with covariates included 
{except for lung cancer}. This, too, is not surprising as the covariates can absorb the differences between neighboring counties and mitigate the residual effects. However, the dependencies among the cancers, the regions and the covariates is complicated and one does not always see a clear pattern. The case for ``smoking'' is pertinent. Figure~\ref{fig: fit_smoking} presents boundaries after accounting for ``smoking''. 
{We see considerably fewer numbers of boundaries for larynx (twenty-two fewer) and esophageal cancer (twenty-five fewer).} The reduction in boundaries in spatial random effects can be attributed to the significant differences between smoking rates in those neighboring counties, i.e. the difference of SIR in neighboring counties is explained by the difference of smoking rates. 
{For example, the smoking rate in Lake is $12.8\%$ higher than that in Sonoma ($25.5\%$ vs. $12.7\%$). Figure~\ref{fig: covariates} reveals that accounting for ``smoking'' eliminates boundaries between the pairs of neighboring counties such as [Lake, Sonoma], [Lake, Mendocino], [Lake, Napa], [Siskiyou, Modoc] and [Shasta, Lassen] for both Larynx and esophageal cancer. The spatial pattern for ``smoking'' in neighboring counties explains most boundaries for larynx and esophageal cancers. At the same time some new boundaries appear after accounting for ``smoking'' such as [Fresno, Monterey], [Fresno, San Benito] for lung cancer, [Monterey, San Luis Obispo] for Larynx cancer and [Del Norte, Humboldt] for colorectal cancer, to offset the difference of smoking rates in pairs of neighboring counties.} It implies the opposite boundary effect of other latent factors against smoking rates in neighboring counties. 
{Figure~\ref{fig: fit_emp} reveals a considerable decrease in difference boundaries for colorectal cancers with more than twenty boundaries fewer after accounting for unemployment. It indicates that the difference boundaries for colorectal cancer are explained by ``unemployment'' in neighboring counties (see Figure~\ref{fig: covariates}). While the number of boundaries detected for the other three cancers mariganally increased in comparison to accounting for ``smoking'' only.} Further discussions about cross-cancer difference boundaries are supplied in Section~\ref{cross-cancer} of the supplementary materials.

\section{Discussion}
The ``MARDP'' detects spatial difference boundaries for multiple correlated diseases that allows us to formulate the problem of areal boundary detection, or ``areal wombling'', as a Bayesian multiple testing problem for spatial random effects. Crucially, the MARDP imposes discrete probability laws on the spatial random effects and we are able to obtain fully model-based estimates of the posterior probabilities for equality of the random effects. This, in turn, allows us to use a Bayesian FDR rule to detect the boundaries.    


Our data analysis on four cancers in California from the SEER database reveals that difference boundaries vary by cancer type under the same FDR threshold. 
{Larynx and Esophageal cancer} exhibits a smoother SIR map with fewer difference boundaries while more are detected for lung cancer. Risk factors also impact difference boundaries for residual spatial random effects for each cancer as accounting for differences in risk factors among neighboring counties can mitigate differences in spatial random effects. These are clearly observed for esophageal, larynx and colorectal cancers, while difference boundaries for lung cancer remain pronounced even after accounting for risk factors. To summarize, the methodology developed here will enable epidemiologists and health policy researchers to identify and hypthesize disparities in health outcomes among neighboring regions and obtain further insights into how and what risk factors or differences in treatment and early detection/diagnosis influence such disparities.

{The proposed methodology will, we hope, generate further explorations into formal statistical inference for difference boundaries. The effectiveness of the FDR, while promising in current demonstrations, should be investigated further in the context of theoretical and empirical implications of different types of multivariate dependencies. The effectiveness of these methods in the context of high-dimensional disease mapping, where dimension can refer to one or all of (a) the number of spatial units; (b) the number of temporal units; and (c) the number of diseases being jointly modeled, should be explored. Finally, we can explore spatial confounding, which occupies a prominent space in disease mapping, in the context of estimating difference boundaries. 
}

\section{Software}\label{sec:software}
Computer programs implementing the numerical examples in the article are available in the public domain at \url{https://github.com/LeiwenG/Multivariate_differenceboundary}.

\section*{Acknowledgements}
The first and second authors were supported in part by grant DMS-1916349 from the Division of Mathematical Sciences (DMS) of the National Science Foundation and by grants R01ES030210 and 5R01ES027027 the National Institute of Environmental Health Sciences (NIEHS). The authors thank the editors and reviewers for their insights and feedback.


\section*{Figures and Tables}

 \begin{table}[H]
   	\centering
   	\caption{Boundary detection results (sensitivity and specificity) in the simulation study (50 datasets generated on the California map) within each disease and across two diseases using MCAR, MDAGAR, CAR$_{ind}$, DAGAR$_{ind}$ and MBLV methods.}\label{tab: rate}
   	\footnotesize
   	\hskip-2cm\begin{tabular}{cccccccccccc}
   		\hline
   		&       & \multicolumn{2}{c}{Disease 1} &   \multicolumn{2}{c}{Disease 2} &     \multicolumn{2}{c}{Disease 1 vs 2} &      &       & \multicolumn{2}{c}{Disease 2 vs 1}   \\
   		\cmidrule(lr){3-4}
   		\cmidrule(lr){5-6}
   		\cmidrule(lr){7-8}
   		\cmidrule(lr){11-12}
   		\multicolumn{1}{l}{$T$} & Methods & \multicolumn{1}{l}{Specificity} & \multicolumn{1}{l}{Sensitivity} & \multicolumn{1}{l}{Specificity} & \multicolumn{1}{l}{Sensitivity} & \multicolumn{1}{l}{Specificity} & \multicolumn{1}{l}{Sensitivity} & \multicolumn{1}{l}{$T$} & Methods & \multicolumn{1}{l}{Specificity} & \multicolumn{1}{l}{Sensitivity} \\
   		\hline
   		60    & MDAGAR & 0.938 & 0.774 & 0.948 & 0.744 & 0.917 & 0.766 & 70    & MDAGAR & 0.915 & 0.730 \\
   		& MCAR  & 0.925 & 0.782 & 0.943 & 0.751 & 0.933 & 0.765 &       & MCAR  & 0.921 & 0.728 \\
   		& DAGAR$_{ind}$ & 0.924 & 0.762 & 0.954 & 0.746 & 0.890 & 0.722 &       & DAGAR$_{ind}$ & 0.888 & 0.712 \\
   		& CAR$_{ind}$ & 0.902 & 0.763 & 0.902 & 0.735 & 0.896 & 0.745 &       & CAR$_{ind}$ & 0.891 & 0.715 \\
   		& MBLV & 0.876 & 0.694 & 0.964 & 0.741 & 0.855 & 0.662 &       & MBLV & 0.865 & 0.674 \\
   		\\
   		65    & MDAGAR & 0.912 & 0.808 & 0.930 & 0.784 & 0.892 & 0.797 & 75    & MDAGAR & 0.889 & 0.761 \\
   		& MCAR  & 0.902 & 0.813 & 0.926 & 0.789 & 0.907 & 0.798 &       & MCAR  & 0.895 & 0.760 \\
   		& DAGAR$_{ind}$ & 0.865 & 0.791 & 0.894 & 0.784 & 0.834 & 0.757 &       & DAGAR$_{ind}$& 0.847 & 0.746 \\
   		& CAR$_{ind}$ & 0.874 & 0.793 & 0.880 & 0.764 & 0.871 & 0.774 &       & CAR$_{ind}$ & 0.862 & 0.744 \\
   		& MBLV & 0.857 & 0.744 & 0.929 & 0.778 & 0.814 & 0.694 &       & MBLV & 0.812 & 0.703 \\
   		\\
   		70    & MDAGAR & 0.881 & 0.843 & 0.900 & 0.820 & 0.856 & 0.823 & 80    & MDAGAR & 0.857 & 0.791 \\
   		& MCAR  & 0.872 & 0.842 & 0.893 & 0.823 & 0.879 & 0.832 &       & MCAR  & 0.869 & 0.793 \\
   		& DAGAR$_{ind}$ & 0.803 & 0.811 & 0.836 & 0.812 & 0.774 & 0.789 &       & DAGAR$_{ind}$ & 0.778 & 0.784 \\
   		& CAR$_{ind}$ & 0.817 & 0.816 & 0.841 & 0.794 & 0.832 & 0.804 &       & CAR$_{ind}$ & 0.813 & 0.774 \\
   		& MBLV & 0.826 & 0.785 & 0.891 & 0.812 & 0.770 & 0.724 &       & MBLV & 0.762 & 0.732 \\
   		\\
   		75    & MDAGAR & 0.838 & 0.869 & 0.857 & 0.850 & 0.814 & 0.848 & 85    & MDAGAR & 0.814 & 0.819 \\
   		& MCAR  & 0.831 & 0.865 & 0.861 & 0.858 & 0.841 & 0.858 &       & MCAR  & 0.834 & 0.824 \\
   		& DAGAR$_{ind}$ & 0.764 & 0.827 & 0.787 & 0.831 & 0.720 & 0.814 &       & DAGAR$_{ind}$ & 0.705 & 0.815 \\
   		& CAR$_{ind}$ & 0.755 & 0.841 & 0.790 & 0.819 & 0.774 & 0.830 &       & CAR$_{ind}$ & 0.737 & 0.810 \\
   		& MBLV & 0.800 & 0.829 & 0.841 & 0.837 & 0.718 & 0.747 &       & MBLV & 0.698 & 0.755 \\
   		\\
   		80    & MDAGAR & 0.782 & 0.885 & 0.807 & 0.875 & 0.766 & 0.868 & 90    & MDAGAR & 0.765 & 0.844 \\
   		& MCAR  & 0.786 & 0.888 & 0.816 & 0.884 & 0.791 & 0.878 &       & MCAR  & 0.793 & 0.856 \\
   		& DAGAR$_{ind}$ & 0.687 & 0.852 & 0.736 & 0.854 & 0.673 & 0.838 &       & DAGAR$_{ind}$ & 0.657 & 0.836 \\
   		& CAR$_{ind}$ & 0.684 & 0.859 & 0.718 & 0.845 & 0.707 & 0.846 &       & CAR$_{ind}$ & 0.681 & 0.833 \\
   		& MBLV & 0.771 & 0.871 & 0.790 & 0.862 & 0.666 & 0.770 &       & MBLV & 0.631 & 0.777 \\
   		\\
   		85    & MDAGAR & 0.712 & 0.903 & 0.751 & 0.895 & 0.715 & 0.888 & 95    & MDAGAR & 0.712 & 0.871 \\
   		& MCAR  & 0.733 & 0.908 & 0.762 & 0.904 & 0.736 & 0.895 &       & MCAR  & 0.739 & 0.881 \\
   		& DAGAR$_{ind}$ & 0.627 & 0.875 & 0.696 & 0.870 & 0.625 & 0.857 &       & DAGAR$_{ind}$ & 0.595 & 0.858 \\
   		& CAR$_{ind}$ & 0.606 & 0.878 & 0.654 & 0.869 & 0.631 & 0.864 &       & CAR$_{ind}$& 0.585 & 0.843 \\
   		& MBLV & 0.732 & 0.905 & 0.737 & 0.884 & 0.609 & 0.789 &       & MBLV & 0.562 & 0.797 \\
   		\hline
   	\end{tabular}
   	\begin{tablenotes}
   		\small
   		\item Note: The first column ``$T$'' is the number of edges fixed as difference boundaries in terms of highest posterior probabilities.
   	\end{tablenotes}
   \end{table}

\begin{table}[H]
	\centering
	\caption{Sensitivity and specificity in the simulation study (50 datasets generated on the California map) for ``disease difference'' in the same region using MCAR, MDAGAR, CAR$_{ind}$, DAGAR$_{ind}$ and MBLV methods.}\label{tab:disease_diff}
	\hskip1cm\begin{tabular}{cccccccc}
		\hline
		$T$ & Methods & Specificity & Sensitivity &$T$ & Methods & Specificity & Sensitivity \\
		\hline
		15    & MDAGAR & 0.912 & 0.676  &20    & MDAGAR & 0.842 & 0.763\\
		& MCAR  & 0.911 & 0.688& & MCAR  & 0.841 & 0.771\\
		& DAGAR$_{ind}$  & 0.889 & 0.596& & DAGAR$_{ind}$  & 0.783 & 0.697\\
		& CAR$_{ind}$ & 0.881 & 0.602 && CAR$_{ind}$  & 0.800 & 0.673 \\
		& MBLV & 0.814 & 0.397& & MBLV & 0.721 & 0.470\\
		\\
		22    & MDAGAR & 0.807 & 0.791& 25& MDAGAR & 0.749 & 0.820 \\
		& MCAR  & 0.789 & 0.796 && MCAR  & 0.735 & 0.828\\
		& DAGAR$_{ind}$  & 0.730 & 0.735 && DAGAR$_{ind}$  & 0.673 & 0.773\\
		& CAR$_{ind}$ & 0.763 & 0.694 && CAR$_{ind}$  & 0.663 & 0.739\\
		& MBLV & 0.681 & 0.493 && MBLV & 0.619 & 0.527 \\
		\\
		30    & MDAGAR & 0.648 & 0.860 \\
		& MCAR  & 0.640 & 0.877 \\
		& DAGAR$_{ind}$  & 0.584 & 0.835 \\
		& CAR$_{ind}$  & 0.555 & 0.795 \\
		& MBLV & 0.507 & 0.563 \\
		\hline
	\end{tabular}
	\begin{tablenotes}
		\small
		\item Note: The first column ``$T$'' is the number of edges fixed as difference boundaries in terms of highest posterior probabilities.
	\end{tablenotes}
\end{table}
 
 
 \begin{footnotesize}
 	\setlength\LTleft{-1.2cm}
 	\setlength\LTright{0pt plus 1fill minus 1fill}
 	\begin{longtable}{ccccc}
 		\caption{Names of adjacent counties that have significant boundary effects from the MDAGAR model for each cancer when $\delta = 0.05$. The numbers in the first column are ranked according to $P(\phi_{id} \neq \phi_{jd} | \bm{y})$. Note: Number $1-35$ for lung cancer, $1-4$ for esophageal cancer and $1-12$ for colorectal cancer are ranked by initial letters with $P(\phi_{id} \neq \phi_{jd} | \bm{y}) = 1$.} \\
 		\hline
 		Rank & Lung (85)  & Esophageal (37) & Layrnx (41) & Colorectal (51)\\
 		\hline
 		1     & Alameda, Contra Costa & \multicolumn{1}{l}{Los Angeles, San Bernardino} & \multicolumn{1}{l}{San Joaquin, Santa Clara} & \multicolumn{1}{l}{Fresno, Monterey} \\
 		2     & Alameda, San Joaquin & \multicolumn{1}{l}{Orange, San Bernardino} & \multicolumn{1}{l}{Santa Clara, Stanislaus} & \multicolumn{1}{l}{Kern, Monterey} \\
 		3     & Alameda, Santa Clara & \multicolumn{1}{l}{Orange, San Diego} & \multicolumn{1}{l}{Kern, Santa Barbara} & \multicolumn{1}{l}{Los Angeles, Orange} \\
 		4     & Alameda, Stanislaus & \multicolumn{1}{l}{San Joaquin, Santa Clara} & \multicolumn{1}{l}{Merced, Santa Clara} & \multicolumn{1}{l}{Los Angeles, San Bernardino} \\
 		5     & Contra Costa, Sacramento & \multicolumn{1}{l}{Los Angeles, Ventura} & \multicolumn{1}{l}{Kern, Ventura} & \multicolumn{1}{l}{Los Angeles, Ventura} \\
 		6     & Contra Costa, San Joaquin & \multicolumn{1}{l}{Orange, Riverside} & \multicolumn{1}{l}{Modoc, Shasta} & \multicolumn{1}{l}{Orange, Riverside} \\
 		7     & Contra Costa, Solano & \multicolumn{1}{l}{Santa Clara, Stanislaus} & \multicolumn{1}{l}{Orange, San Bernardino} & \multicolumn{1}{l}{Orange, San Bernardino} \\
 		8     & Fresno, Monterey & \multicolumn{1}{l}{Modoc, Shasta} & \multicolumn{1}{l}{Orange, Riverside} & \multicolumn{1}{l}{Riverside, San Bernardino} \\
 		9     & Kern, Los Angeles & \multicolumn{1}{l}{Kern, Los Angeles} & \multicolumn{1}{l}{Orange, San Diego} & \multicolumn{1}{l}{Riverside, San Diego} \\
 		10    & Kern, Monterey & \multicolumn{1}{l}{Humboldt, Mendocino} & \multicolumn{1}{l}{Contra Costa, Sacramento} & \multicolumn{1}{l}{San Joaquin, Santa Clara} \\
 		11    & Kern, Santa Barbara & \multicolumn{1}{l}{Lake, Sonoma} & \multicolumn{1}{l}{Alameda, San Joaquin} & \multicolumn{1}{l}{Santa Clara, Stanislaus} \\
 		12    & Kern, Tulare & \multicolumn{1}{l}{Lake, Yolo} & \multicolumn{1}{l}{San Luis Obispo, Santa Barbara} & \multicolumn{1}{l}{Stanislaus, Tuolumne} \\
 		13    & Kern, Ventura & \multicolumn{1}{l}{Lassen, Shasta} & \multicolumn{1}{l}{Placer, Yuba} & \multicolumn{1}{l}{Kern, San Bernardino} \\
 		14    & Lake, Mendocino & \multicolumn{1}{l}{Lake, Mendocino} & \multicolumn{1}{l}{Lassen, Shasta} & \multicolumn{1}{l}{Merced, Santa Clara} \\
 		15    & Lake, Napa & \multicolumn{1}{l}{San Luis Obispo, Santa Barbara} & \multicolumn{1}{l}{Modoc, Siskiyou} & \multicolumn{1}{l}{Placer, Sacramento} \\
 		16    & Lake, Sonoma & \multicolumn{1}{l}{Lake, Napa} & \multicolumn{1}{l}{Lake, Napa} & \multicolumn{1}{l}{Merced, Tuolumne} \\
 		17    & Lake, Yolo & \multicolumn{1}{l}{Alameda, San Joaquin} & \multicolumn{1}{l}{Sierra, Yuba} & \multicolumn{1}{l}{San Francisco, San Mateo} \\
 		18    & Lassen, Shasta & \multicolumn{1}{l}{Modoc, Siskiyou} & \multicolumn{1}{l}{Placer, Sacramento} & \multicolumn{1}{l}{Alameda, Stanislaus} \\
 		19    & Los Angeles, Orange & \multicolumn{1}{l}{Placer, Yuba} & \multicolumn{1}{l}{Lake, Sonoma} & \multicolumn{1}{l}{Marin, Sonoma} \\
 		20    & Los Angeles, San Bernardino & \multicolumn{1}{l}{Mendocino, Tehama} & \multicolumn{1}{l}{Kern, Los Angeles} & \multicolumn{1}{l}{Nevada, Yuba} \\
 		21    & Los Angeles, Ventura & \multicolumn{1}{l}{Sierra, Yuba} & \multicolumn{1}{l}{Nevada, Yuba} & \multicolumn{1}{l}{El Dorado, Sacramento} \\
 		22    & Marin, Sonoma & \multicolumn{1}{l}{Kern, Santa Barbara} & \multicolumn{1}{l}{Los Angeles, San Bernardino} & \multicolumn{1}{l}{Calaveras, Stanislaus} \\
 		23    & Merced, Santa Clara & \multicolumn{1}{l}{Mendocino, Trinity} & \multicolumn{1}{l}{Lake, Yolo} & \multicolumn{1}{l}{Kings, Monterey} \\
 		24    & Modoc, Shasta & \multicolumn{1}{l}{Kern, Ventura} & \multicolumn{1}{l}{Marin, Sonoma} & \multicolumn{1}{l}{Monterey, San Benito} \\
 		25    & Nevada, Yuba & \multicolumn{1}{l}{San Joaquin, Stanislaus} & \multicolumn{1}{l}{Mendocino, Tehama} & \multicolumn{1}{l}{Kern, Santa Barbara} \\
 		26    & Orange, Riverside & \multicolumn{1}{l}{San Francisco, San Mateo} & \multicolumn{1}{l}{Alameda, Santa Clara} & \multicolumn{1}{l}{Nevada, Placer} \\
 		27    & Orange, San Bernardino & \multicolumn{1}{l}{Merced, Santa Clara} & \multicolumn{1}{l}{San Francisco, San Mateo} & \multicolumn{1}{l}{Monterey, San Luis Obispo} \\
 		28    & Orange, San Diego & \multicolumn{1}{l}{Del Norte, Humboldt} & \multicolumn{1}{l}{San Joaquin, Stanislaus} & \multicolumn{1}{l}{Butte, Sutter} \\
 		29    & Placer, Sacramento & \multicolumn{1}{l}{Plumas, Yuba} & \multicolumn{1}{l}{Alameda, Stanislaus} & \multicolumn{1}{l}{Alameda, San Joaquin} \\
 		30    & Placer, Yuba & \multicolumn{1}{l}{Calaveras, Stanislaus} & \multicolumn{1}{l}{Mendocino, Trinity} & \multicolumn{1}{l}{Placer, Yuba} \\
 		31    & San Francisco, San Mateo & \multicolumn{1}{l}{Alameda, Stanislaus} & \multicolumn{1}{l}{El Dorado, Sacramento} & \multicolumn{1}{l}{Modoc, Shasta} \\
 		32    & San Joaquin, Santa Clara & \multicolumn{1}{l}{Alpine, Amador} & \multicolumn{1}{l}{Lake, Mendocino} & \multicolumn{1}{l}{Alpine, Amador} \\
 		33    & San Joaquin, Stanislaus & \multicolumn{1}{l}{Plumas, Shasta} & \multicolumn{1}{l}{Los Angeles, Ventura} & \multicolumn{1}{l}{Sacramento, Sutter} \\
 		34    & San Luis Obispo, Santa Barbara & \multicolumn{1}{l}{Fresno, San Benito} & \multicolumn{1}{l}{Alameda, Contra Costa} & \multicolumn{1}{l}{San Joaquin, Stanislaus} \\
 		35    & Santa Clara, Stanislaus & \multicolumn{1}{l}{Mono, Tuolumne} & \multicolumn{1}{l}{Alpine, Amador} & \multicolumn{1}{l}{Alameda, Contra Costa} \\
 		36    & Monterey, San Luis Obispo & \multicolumn{1}{l}{Plumas, Tehama} & \multicolumn{1}{l}{Humboldt, Mendocino} & \multicolumn{1}{l}{Calaveras, San Joaquin} \\
 		37    & Sacramento, Yolo & \multicolumn{1}{l}{Fresno, Monterey} & \multicolumn{1}{l}{Sacramento, Yolo} & \multicolumn{1}{l}{Butte, Plumas} \\
 		38    & Mendocino, Tehama &       & \multicolumn{1}{l}{Plumas, Yuba} & \multicolumn{1}{l}{Mariposa, Tuolumne} \\
 		39    & Amador, El Dorado &       & \multicolumn{1}{l}{Butte, Plumas} & \multicolumn{1}{l}{Kern, Ventura} \\
 		40    & El Dorado, Sacramento &       & \multicolumn{1}{l}{San Benito, Santa Clara} & \multicolumn{1}{l}{Alameda, Santa Clara} \\
 		41    & Fresno, Tulare &       & \multicolumn{1}{l}{Plumas, Tehama} & \multicolumn{1}{l}{Lassen, Shasta} \\
 		42    & Sierra, Yuba &       &       & \multicolumn{1}{l}{Mariposa, Stanislaus} \\
 		43    & Plumas, Yuba &       &       & \multicolumn{1}{l}{Inyo, San Bernardino} \\
 		44    & Amador, Calaveras &       &       & \multicolumn{1}{l}{Sierra, Yuba} \\
 		45    & Solano, Yolo &       &       & \multicolumn{1}{l}{Inyo, Mono} \\
 		46    & Plumas, Shasta &       &       & \multicolumn{1}{l}{Madera, Merced} \\
 		47    & Plumas, Tehama &       &       & \multicolumn{1}{l}{Orange, San Diego} \\
 		48    & Butte, Plumas &       &       & \multicolumn{1}{l}{Colusa, Glenn} \\
 		49    & Kern, San Bernardino &       &       & \multicolumn{1}{l}{Plumas, Shasta} \\
 		50    & Modoc, Siskiyou &       &       & \multicolumn{1}{l}{Imperial, San Diego} \\
 		51    & Calaveras, San Joaquin &       &       & \multicolumn{1}{l}{Glenn, Mendocino} \\
 		52    & Humboldt, Mendocino &       &       &  \\
 		53    & Napa, Solano &       &       &  \\
 		54    & Inyo, Tulare &       &       &  \\
 		55    & Sutter, Yuba &       &       &  \\
 		56    & Glenn, Mendocino &       &       &  \\
 		57    & Alpine, Amador &       &       &  \\
 		58    & Kern, Kings &       &       &  \\
 		59    & Shasta, Siskiyou &       &       &  \\
 		60    & Kern, San Luis Obispo &       &       &  \\
 		61    & Mendocino, Trinity &       &       &  \\
 		62    & Siskiyou, Trinity &       &       &  \\
 		63    & Placer, Sutter &       &       &  \\
 		64    & Merced, San Benito &       &       &  \\
 		65    & Sutter, Yolo &       &       &  \\
 		66    & Fresno, San Benito &       &       &  \\
 		67    & Butte, Sutter &       &       &  \\
 		68    & Stanislaus, Tuolumne &       &       &  \\
 		69    & Colusa, Lake &       &       &  \\
 		70    & Humboldt, Siskiyou &       &       &  \\
 		71    & Del Norte, Humboldt &       &       &  \\
 		72    & Colusa, Glenn &       &       &  \\
 		73    & Kings, Monterey &       &       &  \\
 		74    & Fresno, Kings &       &       &  \\
 		75    & Calaveras, Stanislaus &       &       &  \\
 		76    & Mendocino, Sonoma &       &       &  \\
 		77    & Inyo, Mono &       &       &  \\
 		78    & Merced, Tuolumne &       &       &  \\
 		79    & Butte, Colusa &       &       &  \\
 		80    & Alpine, Mono &       &       &  \\
 		81    & Lassen, Sierra &       &       &  \\
 		82    & Colusa, Yolo &       &       &  \\
 		83    & Fresno, Mono &       &       &  \\
 		84    & Inyo, Kern &       &       &  \\
 		85    & Madera, Mono &       &       &  \\
 		\hline
 		\label{tab:bound_name}
 	\end{longtable}
 \end{footnotesize}

 \begin{table}[H]
 	\centering
 	\caption{Predictive loss criterion $D$ score under four models: MDAGAR, MCAR, DAGAR$_{ind}$, CAR$_{ind}$ using SEER dataset. The D scores are calculated for each cancer individually and added up to D$_{\text{sum}}$ for all cancers.}\label{tab:D_model}
 	\begin{tabular}{cccccc}
 		\hline
 		Models &D$_{\text{lung}}$ & D$_{\text{esophageal}}$ & D$_{\text{larynx}}$ & D$_{\text{colorectal}}$ & D$_{\text{sum}}$\\
 		\hline
 		MDAGAR & 1.50  & 12.96 & 21.67 & 1.14  & 37.27 \\
 		MCAR  & 1.59  & 13.32 & 21.63 & 1.19  & 37.73\\
 		DAGAR$_{ind}$ & 1.65  & 13.97 & 21.30 & 1.23  & 38.15 \\
 		CAR$_{ind}$ & 1.48  & 12.63 & 21.66& 1.34 & 37.11\\
 		\hline
 	\end{tabular}
 \end{table} 

\begin{table}[H]
	\centering
	\caption{Posterior means (95\% credible intervals) for coefficients and autocorrelation parameters estimated by adding covariates (smoking and unemployed) sequentially}
	\small
	\begin{tabular}{ccccc}
		\hline
		Parameters & Lung  & Esophageal & Larynx & Colorectal \\
		\hline
		Intercept & -0.037 (-0.089, 0.013) & -0.008 (-0.056, 0.047) & -0.052 (-0.112, 0.011) & -0.045 (-0.100, 0.003) \\
		$\rho_d$ & 0.454 (0.286, 0.660)& 0.507 (0.060, 0.870) & 0.615 (0.048, 0.981) & 0.444 (0.041, 0.963)\\
		\\
		Intercept & -0.187 (-0.238, -0.109) & -0.205 (-0.28, -0.061) & -0.250(-0.469, -0.111) & -0.036 (-0.077, -0.004) \\
		Smoking & 0.021 (0.015, 0.028) & 0.03 (0.017, 0.039) & 0.031 (0.019, 0.039) & 0.005 (0.001, 0.009) \\
		$\rho_d$ & 0.400 (0.227, 0.568)& 0.597 (0.242, 0.840) & 0.500 (0.011, 0.980) & 0.554 (0.131, 0.858)\\
		\\
		Intercept & -0.043 (-0.088, -0.023) & -0.051 (-0.209, 0.053) & -0.353 (-0.450, -0.220) & 0.058 (0.029, 0.098) \\
		Smoking & 0.019 (0.015, 0.023) & 0.029 (0.017, 0.043) & 0.027 (0.013, 0.043) & 0.002 (-0.007, 0.09) \\
		Unemployed  & 0.006 (0.000, 0.012) & -0.007 (-0.023, 0.015) & 0.024 (0.002, 0.043) & 0.011 (0.003, 0.022) \\
		$\rho_d$ & 0.229 (0.054, 0.433)& 0.485 (0.032, 0.929) & 0.400 (0.016, 0.887) & 0.696 (0.239, 0.983)\\
		\hline
	\end{tabular}%
	\label{tab:coef}%
\end{table}%

\section*{Supplementary Materials}\label{sec: SM}
\renewcommand{\theequation}{S.\arabic{equation}}
\renewcommand{\thesection}{S.\arabic{section}}
\renewcommand{\thetable}{S.\arabic{table}}
\renewcommand{\thefigure}{S.\arabic{figure}}

\section{Algorithm for MCMC updates}
\label{algorithm}
Algorithm~1 is referenced for model implementation in Section~\ref{implementation}.
\newpage
\noindent{ \rule{\textwidth}{1pt}
	{\fontsize{8}{8}\selectfont
		\textbf{Algorithm~1}: 
		Obtaining posterior inference of $\{\bm{\beta}_d, \bm{\theta}, \bm{\gamma}, \bm{V}, \bm{\tau}, \tau_s, \bm{\rho}, \bm{A}\}$ based on MARDP joint model \\
		\rule{\textwidth}{1pt}
		\begin{enumerate}
			\item update $\bm{\beta}_d | \bm{y}_d, \bm{\phi}_d, \tau_d$
			\begin{align*}
			p(\bm{\beta}_d | \bm{y}_d, \bm{\phi}_d, \tau_d) = N\left(\left(\tau_d\bm{X}_d^\top\bm{X}_d + 1/\sigma_\beta^2\bm{I}_{p_d}\right)^{-1}\tau_d\bm{X}_d^\top(\bm{y}_d-\bm{\phi}_d), \left(\tau_d\bm{X}_d^\top\bm{X}_d + 1/\sigma_\beta^2\bm{I}_{p_d}\right)^{-1}\right)
			\end{align*}
			where $\bm{y}_d = \left(y_{1d}, \dots, y_{nd}\right)^\top$ and $\bm{X}_d = \left(\bm{x}_{1d}, \dots, \bm{x}_{nd}\right)^\top$.
			\item update $\theta_j | \bm{\beta}, \bm{\tau}, \tau_s$, $j = 1, \dots, K$
			\begin{align*}
			p(\theta_j | \bm{\beta}, \bm{\tau}, \tau_s) = N\left(\frac{\sum_{d=1}^q\tau_d\sum_{i:u_{id}=j}\left(y_{id} - \bm{x}_{id}^\top\bm{\beta}_d\right)}{\sum_{d=1}^q\tau_d\sum_{i=1}^nI(u_{id}=j)+\tau_s},\frac{1}{\sum_{d=1}^q\tau_d\sum_{i=1}^nI(u_{id}=j)+\tau_s}\right)
			\end{align*}
			\item update $\gamma_{id} | \bm{\beta},\bm{\theta},\tau_d, \bm{A}, \bm{\rho}$
			\begin{enumerate}
				\item Sample candidate $\gamma_{id}^*$ from $N(\gamma_{id}, s_1^2)$
				\item Compute the corresponding candidate $u_{id}^*$ through $u_{id} = \sum_{j=1}^K j I\left(\sum_{k=1}^{j-1} p_k < F^{(i,d)}(\gamma_{id}) < \sum_{k=1}^{j} p_k \right)$
				\item Accept $\gamma_{id}^*$ with probability
				\begin{align*}
				min\left\{1, \frac{\exp\left(-\frac{1}{2}\bm{\gamma}^{*T}\bm{\Sigma}_\gamma^{-1}\bm{\gamma}^{*}\right)\exp\left(-\frac{\tau_d}{2}\left(y_{id} - \bm{x}_{id}^\top\bm{\beta}_d - \theta_{u_{id}^*}\right)^2\right)}{\exp\left(-\frac{1}{2}\bm{\gamma}^{T}\bm{\Sigma}_\gamma^{-1}\bm{\gamma}\right)\exp\left(-\frac{\tau_d}{2}\left(y_{id} - \bm{x}_{id}^\top\bm{\beta}_d - \theta_{u_{id}}\right)^2\right)}\right\}
				\end{align*}
			\end{enumerate}
			\item update $V_k | \bm{\beta},\bm{\theta},\tau_d, \bm{\gamma}$, $k = 1, \dots, K$
			\begin{enumerate}
				\item Sample candidate $V_k^*$ from $N(V_k, s_2^2)$
				\item Compute the corresponding candidate $\bm{p}^*$ and $\bm{u}^*$, where $\bm{p} = \left\{p_1, \dots, p_K\right\}$ and $\bm{u} = \left\{u_1, \dots, u_N\right\}$
				\item Accept $V_k^*$ with probability
				\begin{align*}
				min\left\{1, \frac{(1-V_k^*)^{\alpha-1}\prod_{d=1}^q\prod_{i=1}^n \exp\left(-\frac{\tau_d}{2}\left(y_{id} - \bm{x}_{id}^\top\bm{\beta}_d - \theta_{u_{id}^*}\right)^2\right)}{(1-V_k)^{\alpha-1}\prod_{d=1}^q\prod_{i=1}^n \exp\left(-\frac{\tau_d}{2}\left(y_{id} - \bm{x}_{id}^\top\bm{\beta}_d - \theta_{u_{id}}\right)^2\right)}\right\}
				\end{align*}
			\end{enumerate}
			\item update $\tau_d |\bm{\beta},\bm{\theta}$
			\begin{align*}
			p(\tau_d | \bm{\beta}, \bm{\theta}) = \Gamma\left(\frac{n}{2} + a_e, \frac{1}{2}\sum_{i=1}^n\left(y_{id} - \bm{x}_{id}^\top\bm{\beta}_d - \theta_{u_{id}}\right)^2 + b_e\right)
			\end{align*}
			\item update $\tau_s |\bm{\theta}$
			\begin{align*}
			p(\tau_s |\bm{\theta}) = \Gamma\left(\frac{K}{2} + a_s, \frac{1}{2}\sum_{j=1}^K\theta_j^2 +b_s\right)
			\end{align*}
			\item update $\bm{\rho} |\bm{\gamma}$
			\begin{enumerate}
				\item  Let $\bm{\eta} = \mbox{logit}(\bm{\rho})$ and sample the candidate $\eta_d^*$ from $N(\eta_d, s_3^2)$, then $\rho_d^* = \frac{\exp(\eta_d^*)}{1+\exp(\eta_d^*)}$
				\item Accept $\bm{\rho}^*$ with probability
				\begin{align*}
				min\left\{1, \frac{|\bm{\Sigma}_\gamma^*|^{-\frac{N}{2}}\exp\left(-\frac{1}{2}\bm{\gamma}^{T}\bm{\Sigma}_\gamma^{*-1}\bm{\gamma}\right)\prod_{d=1}^q\rho_d^*(1-\rho_d^*)}{|\bm{\Sigma}_\gamma|^{-\frac{N}{2}}\exp\left(-\frac{1}{2}\bm{\gamma}^{T}\bm{\Sigma}_\gamma^{-1}\bm{\gamma}\right)\prod_{d=1}^q\rho_d(1-\rho_d)}\right\}
				\end{align*}
			\end{enumerate}
			\item update $\bm{A} |\bm{\gamma}$
			\begin{enumerate}
				\item Let $z_{dd} = log(a_{dd})$ and sample candidates $z_{dd}^*$ from $N(z_{dd}, s_4^2)$
				\item For off-diagonal elements $a_{dh}, d \neq h$, $a_{dh}^*$ are sampled from $N(a_{dh}, s_5^2)$
				\item Accept $\bm{A}^*$ with probability
				\begin{align*}
				min\left\{1, \frac{|\bm{\Sigma}_\gamma^*|^{-\frac{N}{2}}exp\left(-\frac{1}{2}\bm{\gamma}^{T}\bm{\Sigma}_\gamma^{*-1}\bm{\gamma}\right)p(\bm{A}^*)\prod_{d=1}^qa_{dd}^*}{|\bm{\Sigma}_\gamma|^{-\frac{N}{2}}exp\left(-\frac{1}{2}\bm{\gamma}^{T}\bm{\Sigma}_\gamma^{-1}\bm{\gamma}\right)p(\bm{A})\prod_{d=1}^qa_{dd}}\right\}
				\end{align*}
			\end{enumerate}
		\end{enumerate}	
		\vspace*{-8pt}
		\rule{\textwidth}{1pt}
} }
\section{Evaluation of parameter estimates in simulation study}
\label{sup_sim}
For the simulation study in Section~\ref{sim}, we evaluated parameter estimates from MAGAR, MCAR, DAGAR$_{ind}$ and CAR$_{ind}$ models. Table~\ref{tab:cp} shows the 
{average of the posterior means over 50 data sets along with the averages of the lower quantile and the upper quantile as a summary. We also present the coverage probability (CP) defined as the proportion of data sets where the $95\%$ credible intervals included the true parameter values out of the $50$ datasets. All the models appear to provide effective coverages between $90\% - 100\%$ for slope parameters $\bm{\beta}_1 = (\beta_{11}, \beta_{12})^\top$ and $\bm{\beta}_2 = (\beta_{21}, \beta_{22})^\top$.} In terms of the precision parameters for random noise, $\tau_1$ and $\tau_2$, MDAGAR and MCAR offer comparable coverages at around $85\%$, while the two independent-disease models present much lower coverage probabilities as they fail to acquire dependent spatial structures for random effects. 

The $95\%$ credible intervals for the spatial autocorrelation parameters $\rho_1$ and $\rho_2$ estimated from CAR-based models are wide (nearly covering the entire interval $(0, 1)$). Therefore, we computed the mean squared errors (MSE) (measuring the error between estimated values and the true values) over 
{$50$} datasets instead. Table~\ref{tab:MSE} shows estimated MSEs of $\rho_1$ and $\rho_2$ from each model. Recall that the true values $\rho_1 = 0.2$ and $\rho_2 = 0.8$. Unsurprisingly, MDAGAR delivers better inferential performance for $\rho_1$, while MCAR is superior for $\rho_2$. This finding is consistent with \citet{datta2018spatial} who report that (univariate) DAGAR delivers better estimates of the autocorrelation parameters when $\rho$ is not too high. 

\section{Impact of covariates on mutual cross-cancer difference boundaries}
\label{cross-cancer}
\begin{figure}[H]
	\centering
	\centering
	\includegraphics[width=150mm]{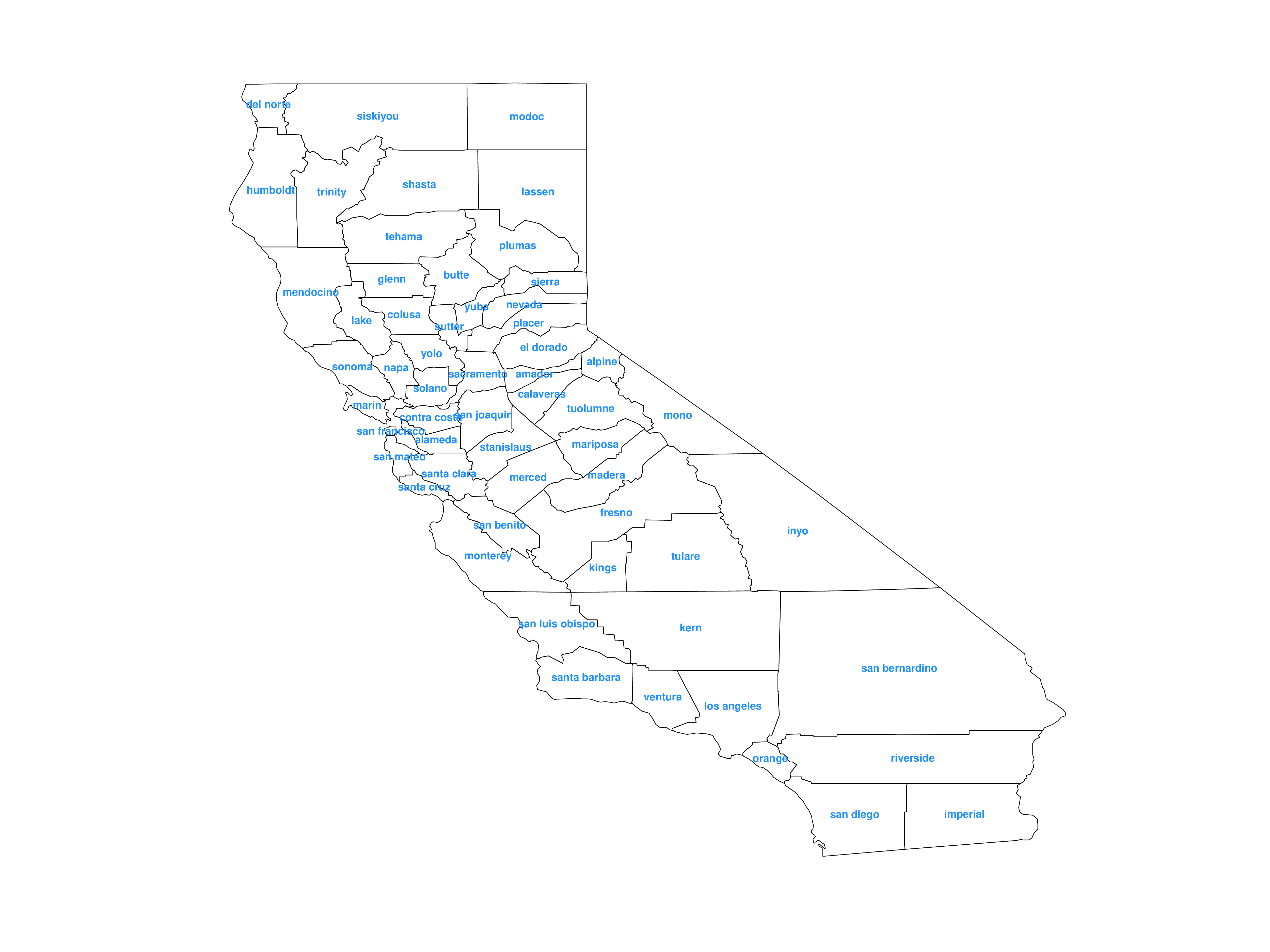}
	\caption{California map with county names labeled}\label{fig: county_map}
\end{figure}

\begin{figure}[H]
	\centering
	\begin{subfigure}{0.5\linewidth}
		\centering
		\includegraphics[width=85mm]{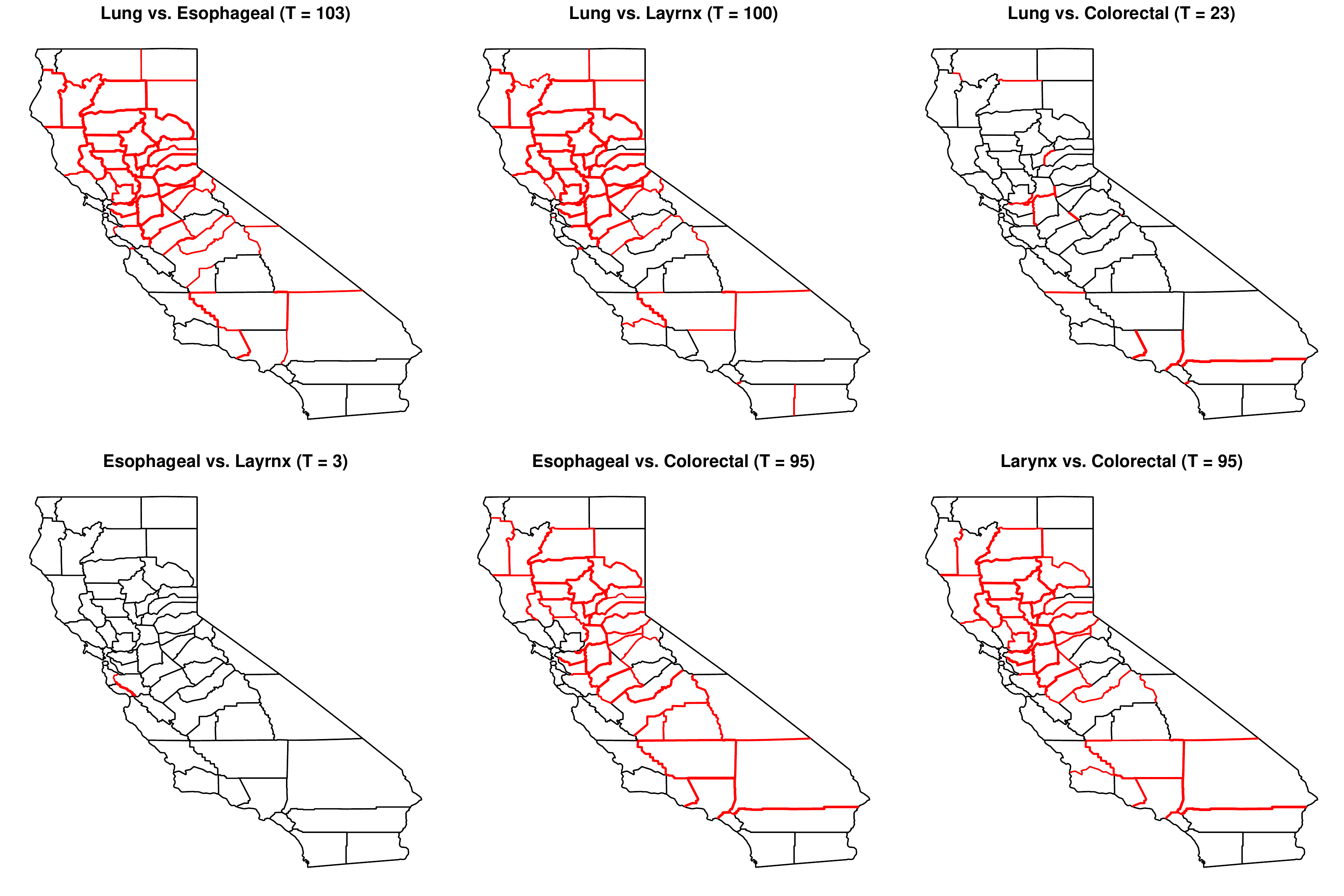}
		\caption{Smoking}\label{fig: cross_smoking}
	\end{subfigure}
	\vfill
	\vskip 1cm
	\begin{subfigure}{0.5\linewidth}
		\centering
		\includegraphics[width=85mm]{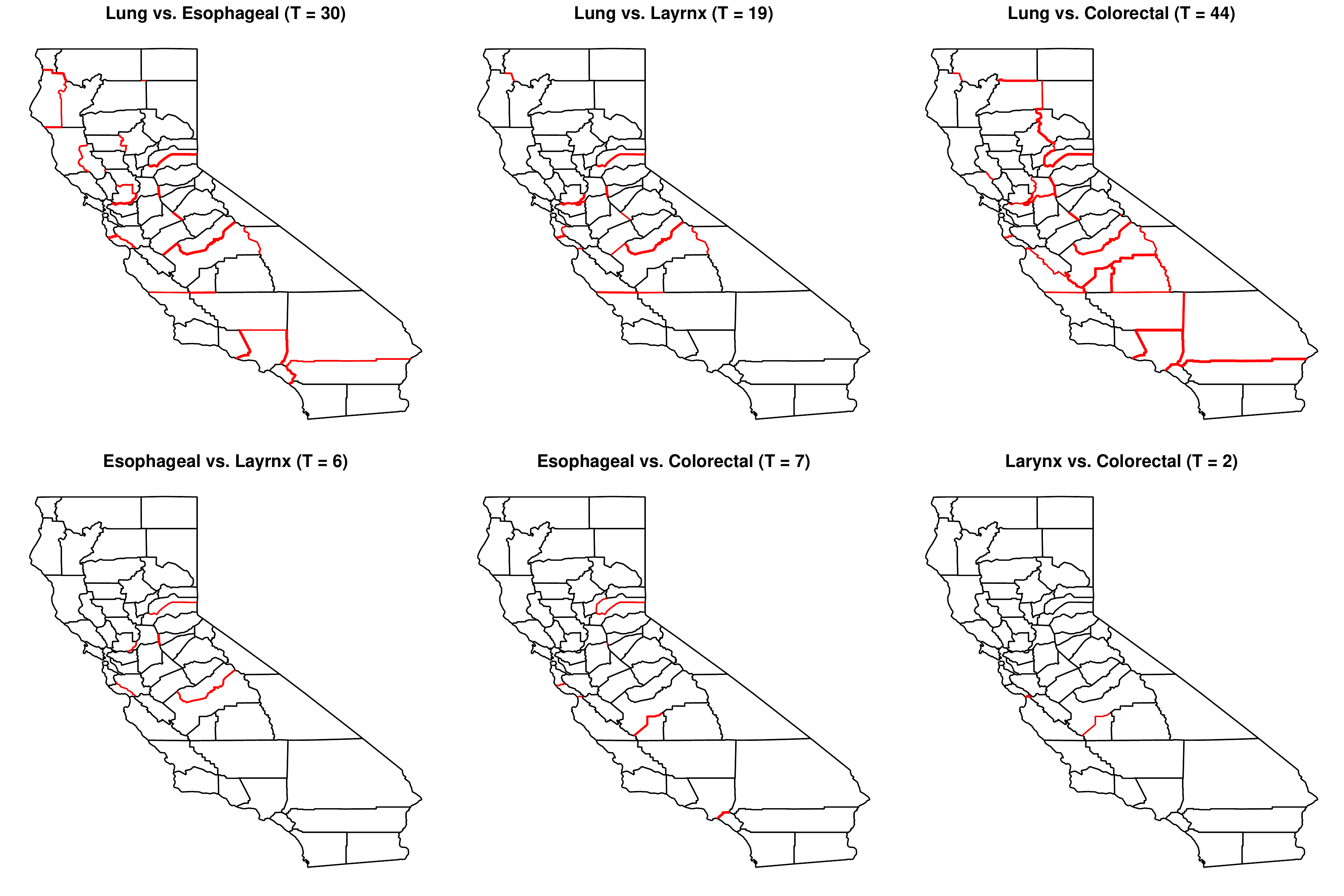}
		\caption{Smoking, Unemployed}\label{fig: cross_emp}
	\end{subfigure}
	\caption{Mutual cross-cancer difference boundaries (highlighted in red) detected by MDAGAR for each pair of cancers after accounting for (a) smoking and (b) smoking and unemployed when $\delta = 0.05$. The values in brackets are the number of difference boundaries detected.}\label{fig: cross_diff_cov}
\end{figure}

Figure~\ref{fig: county_map} presents a map of California with the names and boundaries of each county. Accounting for covariates also affects the detection of mutual cross-cancer difference boundaries for each pair of cancers. Figure~\ref{fig: cross_diff_cov} shows mutual cross-cancer difference boundaries detected for each pair of cancers after accounting for only ``smoking'' in \ref{fig: cross_smoking} and accounting for  both``smoking'' and ``unemployed'' in \ref{fig: cross_emp} when $\delta = 0.05$. 
{Accounting only for ``smoking'' eliminates the mutual cross-cancer difference boundaries for pairs of [lung, colorectal] and [esophageal, larynx] while increasing boundaries for the other pairs. From the individual cancer analysis as discussed in Section \ref{analysis}, the spatial pattern of ``smoking'' mitigates the difference boundaries for esophageal and larynx cancers but not so for lung and colorectal cancers. In fact, accounting for smoking may elicit greater heterogeneity in the spatial distribution rates of lung and colorectal cancers	
	and of esophageal and larynx cancers, thereby introducing more cross-cancer difference boundaries. Most of the mutual cross-cancer difference boundaries are explained after accounting for ``unemployment'', especially across esophageal, larynx and colorectal cancers where only very few boundaries between pairwise cancers are evinced from the residual spatial random effects.}

\newpage

\begin{table}[H]
	\centering
	\caption{Average of posterior means (average lower quantile, average upper quantile) over 50 data sets and coverage probability ($\%$) of true parameter values estimated from MDAGAR, MCAR, DAGAR$_{ind}$ and CAR$_{ind}$}
	\small
	\hspace*{-1.5cm}	\begin{tabular}{cccccccc}
		\hline
		& $\beta_{11}$ & $\beta_{12}$ & $\beta_{21}$ & $\beta_{22}$ & $\tau_{1}$ & $\tau_{2}$  \\
		\hline
		MDAGAR & 2.14 (1.31, 3.04)   & 5.00 (4.89, 5.11)  & 0.94 (0.10, 1.83)   & 6.00 (5.89, 6.11)& 9.33 (5.43, 14.96) & 9.31 (5.42, 15.02)\\
		(\%)& 98  & 90  & 100   & 96  & 82  & 86  \\
		MCAR  & 2.18 (1.30, 3.06)  & 4.99 (4.89, 5.10)  & 0.97 (0.09, 1.86)& 6.00 (5.89, 6.11)    & 9.63 (5.63, 15.52) & 9.85 (5.67, 15.75)\\
		(\%)& 96.0  & 94.0  & 98 & 92.0    & 92    & 82.0 \\
		DAGAR$_{ind}$ & 2.27 (1.02, 3.55)   & 4.99 (4.85, 5.14)  & 1.14	(-0.11, 2.41)   & 6.00 (5.84, 6.16)& 5.92 (3.39, 9.23) & 5.76 (3.25, 9.16) \\
		(\%)& 100   & 89.7  & 100   & 98.7  & 40.7   & 22.7   \\
		CAR$_{ind}$ & 2.33 (1.40, 3.27)   & 4.99 (4.87, 5.12)  & 1.13 (0.20, 2.07)  & 5.99 (5.86, 6.12)  & 6.73 (4.01, 10.29) & 6.72 (3.83, 10.85)  \\
		(\%)& 100   & 94.0  & 100   & 96.0  & 50.0     & 58.0  \\
		\hline
	\end{tabular}%
	\label{tab:cp}%
\end{table}%

\begin{table}[H]
	\centering
	\caption{Estimated MSEs of autocorrelation parameters $\rho_1$ and $\rho_2$ estimated from MAGAR, MCAR, DAGAR$_{ind}$ and CAR$_{ind}$.}
	\begin{tabular}{ccc}
		\hline
		Methods & MSE$_{\rho_1}$  & MSE$_{\rho_2}$ \\
		\hline
		MDAGAR & 0.029  & 0.169 \\
		MCAR  & 0.149 & 0.081 \\
		DAGAR$_{ind}$ & 0.589 & 0.028 \\
		CAR$_{ind}$ & 0.022 & 0.193 \\
		\hline
		\hline
	\end{tabular}%
	\label{tab:MSE}%
\end{table}%

\bibliographystyle{biorefs}
\bibliography{Banerjee}

\newcommand{\noop}[1]{}
\begin{thebibliography}{99}

\bibitem[Agrawal \emph{and others}(2018)Agrawal, Markert and
  Agrawal]{agrawal2018risk}
\textsc{Agrawal, Kriti, Markert, Ronald~J and Agrawal, Sangeeta}. (2018).
\newblock Risk factors for adenocarcinoma and squamous cell carcinoma of the
  esophagus and lung.
\newblock {\em Hypertension\/}~\textbf{61}(46), 0--09.

\bibitem[Akhtar \emph{and others}(2010)Akhtar, Bhargava, Shameem, Singh,
  Baneen, Khan, Hassan and Sharma]{akhtar2010second}
\textsc{Akhtar, Jamal, Bhargava, Rakesh, Shameem, Mohammad, Singh, Saurabh~K,
  Baneen, Ummul, Khan, Nafees~Ahmad, Hassan, Jassem and Sharma, Prakhar}.
  (2010).
\newblock Second primary lung cancer with glottic laryngeal cancer as index
  tumor--a case report.
\newblock {\em Case Reports in Oncology\/}~\textbf{3}(1), 35--39.

\bibitem[Banerjee \emph{and others}(2014)Banerjee, Carlin and
  Gelfand]{banerjee2014hierarchical}
\textsc{Banerjee, Sudipto, Carlin, Bradley~P and Gelfand, Alan~E}. (2014).
\newblock {\em Hierarchical Modeling and Analysis for Spatial Data\/}. CRC
  Press, Boca Raton, FL.

\bibitem[Banerjee and Gelfand(2006)Banerjee and Gelfand]{banerjee2006bayesian}
\textsc{Banerjee, Sudipto and Gelfand, Alan~E}. (2006).
\newblock Bayesian wombling: Curvilinear gradient assessment under spatial
  process models.
\newblock {\em Journal of the American Statistical
  Association\/}~\textbf{101}(476), 1487--1501.

\bibitem[Benjamini and Hochberg(1995)Benjamini and
  Hochberg]{benjamini1995controlling}
\textsc{Benjamini, Yoav and Hochberg, Yosef}. (1995).
\newblock Controlling the false discovery rate: A practical and powerful
  approach to multiple testing.
\newblock {\em Journal of the Royal statistical society: series B
  (Methodological)\/}~\textbf{57}(1), 289--300.

\bibitem[Berchuck \emph{and others}(2019)Berchuck, Mwanza and
  Warren]{berchuck2019jasa}
\textsc{Berchuck, Samuel~I., Mwanza, Jean-Claude and Warren, Joshua~L.} (2019).
\newblock Diagnosing glaucoma progression with visual field data using a
  spatiotemporal boundary detection method.
\newblock {\em Journal of the American Statistical
  Association\/}~\textbf{114}(527), 1063--1074.
\newblock PMID: 31662589.

\bibitem[Besag(1974)Besag]{besag1974spatial}
\textsc{Besag, Julian}. (1974).
\newblock Spatial interaction and the statistical analysis of lattice systems.
\newblock {\em Journal of the Royal Statistical Society: Series B
  (Methodological)\/}~\textbf{36}(2), 192--225.

\bibitem[Besag \emph{and others}(1991)Besag, York and
  Molli{\'e}]{besag1991bayesian}
\textsc{Besag, Julian, York, Jeremy and Molli{\'e}, Annie}. (1991).
\newblock Bayesian image restoration, with two applications in spatial
  statistics.
\newblock {\em Annals of the Institute of Statistical
  Mathematics\/}~\textbf{43}(1), 1--20.

\bibitem[Bradley \emph{and others}(2018)Bradley, Holan and
  Wikle]{bradley2018computationally}
\textsc{Bradley, Jonathan~R, Holan, Scott~H and Wikle, Christopher~K}. (2018).
\newblock Computationally efficient multivariate spatio-temporal models for
  high-dimensional count-valued data (with discussion).
\newblock {\em Bayesian Analysis\/}~\textbf{13}(1), 253--310.

\bibitem[Bradley \emph{and others}(2015)Bradley, Holan, Wikle
  et~al.]{bradley2015multivariate}
\textsc{Bradley, Jonathan~R, Holan, Scott~H, Wikle, Christopher~K  \emph{and
  others}}. (2015).
\newblock Multivariate spatio-temporal models for high-dimensional areal data
  with application to longitudinal employer-household dynamics.
\newblock {\em The Annals of Applied Statistics\/}~\textbf{9}(4), 1761--1791.

\bibitem[{California Department of Public Health, California Tobacco Control
  Program}(2018){California Department of Public Health, California Tobacco
  Control Program}]{california2018california}
\textsc{{California Department of Public Health, California Tobacco Control
  Program}}. (2018).
\newblock {\em California Tobacco Facts and Figures 2018\/}. Sacramento, CA:
  California Department of Public Health.

\bibitem[Carlin and Ma(2007)Carlin and Ma]{carlin2007bayesian}
\textsc{Carlin, Bradley~P and Ma, Haijun}. (2007).
\newblock Bayesian multivariate areal wombling for multiple disease boundary
  analysis.
\newblock {\em Bayesian Analysis\/}~\textbf{2}(2), 281--302.

\bibitem[Corpas-Burgos and Martinez-Beneito(2020)Corpas-Burgos and
  Martinez-Beneito]{corpas-burgos2020serra}
\textsc{Corpas-Burgos, Francisca and Martinez-Beneito, Miguel~A.} (2020).
\newblock On the use of adaptive spatial weight matrices from disease mapping
  multivariate analyses.
\newblock {\em Stochastic Environmental Research and Risk
  Assessment\/}~\textbf{34}, 531–544.

\bibitem[Datta \emph{and others}(2019)Datta, Banerjee, Hodges and
  Gao]{datta2018spatial}
\textsc{Datta, Abhirup, Banerjee, Sudipto, Hodges, James~S. and Gao, Leiwen}.
  (2019).
\newblock {Spatial Disease Mapping Using Directed Acyclic Graph Auto-Regressive
  (DAGAR) Models}.
\newblock {\em Bayesian Analysis\/}~\textbf{14}(4), 1221 -- 1244.

\bibitem[Gamerman and Lopes(2006)Gamerman and Lopes]{gamerman2006markov}
\textsc{Gamerman, Dani and Lopes, Hedibert~F}. (2006).
\newblock {\em Markov Chain Monte Carlo: Stochastic Simulation for Bayesian
  Inference\/}. Chapman and Hall/CRC.

\bibitem[Gao \emph{and others}(2022)Gao, Datta and
  Banerjee]{gao2021hierarchical}
\textsc{Gao, Leiwen, Datta, Abhirup and Banerjee, Sudipto}. (2022).
\newblock Hierarchical multivariate directed acyclic graph auto-regressive
  ({MDAGAR}) models for spatial diseases mapping.
\newblock {\em Statistics in Medicine\/}.

\bibitem[Gelfand and Ghosh(1998)Gelfand and Ghosh]{gelfand1998model}
\textsc{Gelfand, Alan~E and Ghosh, Sujit~K}. (1998).
\newblock Model choice: A minimum posterior predictive loss approach.
\newblock {\em Biometrika\/}~\textbf{85}(1), 1--11.

\bibitem[Gelfand and Vounatsou(2003)Gelfand and Vounatsou]{gelfand2003proper}
\textsc{Gelfand, Alan~E and Vounatsou, Penelope}. (2003).
\newblock Proper multivariate conditional autoregressive models for spatial
  data analysis.
\newblock {\em Biostatistics\/}~\textbf{4}(1), 11--15.

\bibitem[Hanson \emph{and others}(2015)Hanson, Banerjee, Li and
  McBean]{hanson2015spatial}
\textsc{Hanson, Timothy, Banerjee, Sudipto, Li, Pei and McBean, Alexander}.
  (2015).
\newblock Spatial boundary detection for areal counts.
\newblock In:  {\em Nonparametric Bayesian Inference in Biostatistics\/}.
  Springer, pp.\  377--399.

\bibitem[Jacquez and Greiling(2003)Jacquez and Greiling]{jacquez2003geographic}
\textsc{Jacquez, Geoffrey~M and Greiling, Dunrie~A}. (2003{\em a}).
\newblock Geographic boundaries in breast, lung and colorectal cancers in
  relation to exposure to air toxics in long island, new york.
\newblock {\em International Journal of Health Geographics\/}~\textbf{2}(1),
  1--22.

\bibitem[Jacquez and Greiling(2003)Jacquez and Greiling]{jacquez2003local}
\textsc{Jacquez, Geoffrey~M and Greiling, Dunrie~A}. (2003{\em b}).
\newblock Local clustering in breast, lung and colorectal cancer in long
  island, new york.
\newblock {\em International Journal of Health Geographics\/}~\textbf{2}(1),
  1--12.

\bibitem[Jin \emph{and others}(2007)Jin, Banerjee and Carlin]{jin2007order}
\textsc{Jin, Xiaoping, Banerjee, Sudipto and Carlin, Bradley~P}. (2007).
\newblock Order-free co-regionalized areal data models with application to
  multiple-disease mapping.
\newblock {\em Journal of the Royal Statistical Society: Series B (Statistical
  Methodology)\/}~\textbf{69}(5), 817--838.

\bibitem[Jin \emph{and others}(2005)Jin, Carlin and
  Banerjee]{jin2005generalized}
\textsc{Jin, Xiaoping, Carlin, Bradley~P and Banerjee, Sudipto}. (2005).
\newblock Generalized hierarchical multivariate car models for areal data.
\newblock {\em Biometrics\/}~\textbf{61}(4), 950--961.

\bibitem[Kissling and Carl(2008)Kissling and Carl]{kissling2008spatial}
\textsc{Kissling, W~Daniel and Carl, Gudrun}. (2008).
\newblock Spatial autocorrelation and the selection of simultaneous
  autoregressive models.
\newblock {\em Global Ecology and Biogeography\/}~\textbf{17}(1), 59--71.

\bibitem[Koch(2005)Koch]{koch2005cartographies}
\textsc{Koch, Tom}. (2005).
\newblock {\em Cartographies of disease: maps, mapping, and medicine\/}. Esri
  Press Redlands, CA.

\bibitem[Kurishima \emph{and others}(2018)Kurishima, Miyazaki, Watanabe,
  Shiozawa, Ishikawa, Satoh and Hizawa]{kurishima2018lung}
\textsc{Kurishima, Koich, Miyazaki, Kunihiko, Watanabe, Hiroko, Shiozawa,
  Toshihiro, Ishikawa, Hiroichi, Satoh, Hiroaki and Hizawa, Nobuyuki}. (2018).
\newblock Lung cancer patients with synchronous colon cancer.
\newblock {\em Molecular and Clinical Oncology\/}~\textbf{8}(1), 137--140.

\bibitem[Lawson \emph{and others}(2016)Lawson, Banerjee, Haining and
  Ugarte]{lawson2016handbook}
\textsc{Lawson, B. Andrew, Banerjee, Sudipto, Haining, Robert and Ugarte, D.
  Maria}. (2016).
\newblock {\em Handbook of Spatial Epidemiology\/}. CRC press, Boca Raton, FL.

\bibitem[Lee \emph{and others}(2021)Lee, Meeks and Pettersson]{lee2021statcomp}
\textsc{Lee, Duncan, Meeks, Kitty and Pettersson, William}. (2021).
\newblock Improved inference for areal unit count data using graph-based
  optimisation.
\newblock {\em Stat. Comput.\/}~\textbf{31}(4), 51.

\bibitem[Li \emph{and others}(2012)Li, Banerjee, Carlin and
  McBean]{banerjee2012bayesian}
\textsc{Li, Pei, Banerjee, Sudipto, Carlin, Bradley~P and McBean, Alexander~M}.
  (2012).
\newblock Bayesian areal wombling using false discovery rates.
\newblock {\em Statistics and its Interface\/}~\textbf{5}(2), 149--158.

\bibitem[Li \emph{and others}(2015)Li, Banerjee, Hanson and
  McBean]{li2015bayesian}
\textsc{Li, Pei, Banerjee, Sudipto, Hanson, Timothy~A and McBean, Alexander~M}.
  (2015).
\newblock Bayesian models for detecting difference boundaries in areal data.
\newblock {\em Statistica Sinica\/}, 385--402.

\bibitem[Li \emph{and others}(2011)Li, Banerjee and McBean]{li2011mining}
\textsc{Li, P, Banerjee, S and McBean, AM}. (2011).
\newblock Mining edge effects in areally referenced spatial data: A bayesian
  model choice approach.
\newblock {\em Geoinformatica\/}~\textbf{15}, 435--454.

\bibitem[Lindstr{\"o}m \emph{and others}(2017)Lindstr{\"o}m, Finucane,
  Bulik-Sullivan, Schumacher, Amos, Hung, Rand, Gruber, Conti, Permuth
  et~al.]{lindstrom2017quantifying}
\textsc{Lindstr{\"o}m, Sara, Finucane, Hilary, Bulik-Sullivan, Brendan,
  Schumacher, Fredrick~R, Amos, Christopher~I, Hung, Rayjean~J, Rand, Kristin,
  Gruber, Stephen~B, Conti, David, Permuth, Jennifer~B  \emph{and others}}.
  (2017).
\newblock Quantifying the genetic correlation between multiple cancer types.
\newblock {\em Cancer Epidemiology and Prevention Biomarkers\/}~\textbf{26}(9),
  1427--1435.

\bibitem[Lu and Carlin(2005)Lu and Carlin]{lu2005bayesian}
\textsc{Lu, Haolan and Carlin, Bradley~P}. (2005).
\newblock Bayesian areal wombling for geographical boundary analysis.
\newblock {\em Geographical Analysis\/}~\textbf{37}(3), 265--285.

\bibitem[Lu \emph{and others}(2007)Lu, Reilly, Banerjee and
  Carlin]{lu2007bayesian}
\textsc{Lu, Haolan, Reilly, Cavan~S, Banerjee, Sudipto and Carlin, Bradley~P}.
  (2007).
\newblock Bayesian areal wombling via adjacency modeling.
\newblock {\em Environmental and ecological statistics\/}~\textbf{14}(4),
  433--452.

\bibitem[Ma \emph{and others}(2010)Ma, Carlin and Banerjee]{ma2010hierarchical}
\textsc{Ma, Haijun, Carlin, Bradley~P and Banerjee, Sudipto}. (2010).
\newblock Hierarchical and joint site-edge methods for medicare hospice service
  region boundary analysis.
\newblock {\em Biometrics\/}~\textbf{66}(2), 355--364.

\bibitem[MacNab(2016)MacNab]{mcnab2016}
\textsc{MacNab, Ying~C.} (2016).
\newblock Linear models of coregionalization for multivariate lattice data: a
  general framework for coregionalized multivariate car models.
\newblock {\em Statistics in Medicine\/}~\textbf{35}(21), 3827--3850.

\bibitem[MacNab(2018)MacNab]{mcnab2018}
\textsc{MacNab, Ying~C.} (2018, September).
\newblock {Some recent work on multivariate Gaussian Markov random fields (with
  discussion)}.
\newblock {\em TEST: An Official Journal of the Spanish Society of Statistics
  and Operations Research\/}~\textbf{27}(3), 497--541.

\bibitem[Mardia(1988)Mardia]{mardia1988multi}
\textsc{Mardia, KV}. (1988).
\newblock Multi-dimensional multivariate gaussian markov random fields with
  application to image processing.
\newblock {\em Journal of Multivariate Analysis\/}~\textbf{24}(2), 265--284.

\bibitem[M{\"u}ller \emph{and others}(2004)M{\"u}ller, Parmigiani, Robert and
  Rousseau]{muller2004optimal}
\textsc{M{\"u}ller, Peter, Parmigiani, Giovanni, Robert, Christian and
  Rousseau, Judith}. (2004).
\newblock Optimal sample size for multiple testing: The case of gene expression
  microarrays.
\newblock {\em Journal of the American Statistical
  Association\/}~\textbf{99}(468), 990--1001.

\bibitem[{National Cancer Institute}(2019){National Cancer Institute}]{seer}
\textsc{{National Cancer Institute}}. (2019, Aug).
\newblock Seer*stat software.

\bibitem[Perone~Pacifico \emph{and others}(2004)Perone~Pacifico, Genovese,
  Verdinelli and Wasserman]{perone2004false}
\textsc{Perone~Pacifico, Marco, Genovese, Christopher, Verdinelli, Isabella and
  Wasserman, Larry}. (2004).
\newblock False discovery control for random fields.
\newblock {\em Journal of the American Statistical
  Association\/}~\textbf{99}(468), 1002--1014.

\bibitem[Qu \emph{and others}(2021)Qu, Bradley and Niu]{qu2021boundary}
\textsc{Qu, Kai, Bradley, Jonathan~R and Niu, Xufeng}. (2021).
\newblock Boundary detection using a bayesian hierarchical model for multiscale
  spatial data.
\newblock {\em Technometrics\/}~\textbf{63}(1), 64--76.

\bibitem[Rue and Held(2005)Rue and Held]{rueheld04}
\textsc{Rue, Havard and Held, Leonard}. (2005).
\newblock {\em Gaussian Markov Random Fields : Theory and Applications\/},
  Monographs on statistics and applied probability. Chapman and Hall/CRC Press,
  Boca Raton, FL.

\bibitem[Rushworth \emph{and others}(2017)Rushworth, Lee and
  Sarran]{rushworth2017jrssc}
\textsc{Rushworth, Alastair, Lee, Duncan and Sarran, Christophe}. (2017,
  January).
\newblock {An adaptive spatiotemporal smoothing model for estimating trends and
  step changes in disease risk}.
\newblock {\em Journal of the Royal Statistical Society Series
  C\/}~\textbf{66}(1), 141--157.

\bibitem[Sain and Cressie(2007)Sain and Cressie]{sain2007}
\textsc{Sain, Stephan~R. and Cressie, Noel}. (2007, September).
\newblock {A spatial model for multivariate lattice data}.
\newblock {\em Journal of Econometrics\/}~\textbf{140}(1), 226--259.

\bibitem[Santafé \emph{and others}(2021)Santafé, Adin, Lee and
  Ugarte]{santafe2021smmr}
\textsc{Santafé, Guzman, Adin, Aritz, Lee, Duncan and Ugarte,
  Maŕıa~Dolores}. (2021).
\newblock Dealing with risk discontinuities to estimate cancer mortality risks
  when the number of small areas is large.
\newblock {\em Statistical Methods in Medical Research\/}~\textbf{30}(1),
  6--21.
\newblock PMID: 33595401.

\bibitem[Sethuraman(1994)Sethuraman]{sethuraman1994constructive}
\textsc{Sethuraman, Jayaram}. (1994).
\newblock A constructive definition of dirichlet priors.
\newblock {\em Statistica sinica\/}, 639--650.

\bibitem[Shi and Chen(2004)Shi and Chen]{shi2004frequencies}
\textsc{Shi, Weixing and Chen, Shuqing}. (2004).
\newblock Frequencies of poor metabolizers of cytochrome p450 2c19 in esophagus
  cancer, stomach cancer, lung cancer and bladder cancer in chinese population.
\newblock {\em World Journal of Gastroenterology: WJG\/}~\textbf{10}(13), 1961.

\bibitem[Tansey \emph{and others}(2018)Tansey, Koyejo, Poldrack and
  Scott]{tansey2018false}
\textsc{Tansey, Wesley, Koyejo, Oluwasanmi, Poldrack, Russell~A and Scott,
  James~G}. (2018).
\newblock False discovery rate smoothing.
\newblock {\em Journal of the American Statistical
  Association\/}~\textbf{113}(523), 1156--1171.

\bibitem[Womble(1951)Womble]{womble1951differential}
\textsc{Womble, William~H}. (1951).
\newblock Differential systematics.
\newblock {\em Science\/}~\textbf{114}(2961), 315--322.

\end{thebibliography}

\end{document}